\documentclass{sig-alternate-05-2015}

	\usepackage[draft]{todonotes}   
	\usepackage[capitalise,nameinlink]{cleveref}
	\usepackage{xspace}

	\newcommand{\hbi}{{\textit{History-based Insights}}\xspace}
	
	\newcommand{\HB}{{\textit{HB}}\xspace}
	\newcommand{\BL}{{\textit{BL}}\xspace}
	
	\newcommand{\eg}{e.\,g.,\ }
	\newcommand{\ie}{i.\,e.,\ }
	\newcommand{\VFC}{\mbox{\textit{VFC}}}
	
	\newcommand{\FA}{\mbox{\textit{FA}}}
	\newcommand{\EHB}{\mbox{\textit{EHB}}}
	\newcommand{\EBL}{\mbox{\textit{EBL}}}
	\usepackage{tikz} \newcommand*\circled[1]{\tikz[baseline=(char.base)]{ \node[shape=circle,draw,inner sep=1pt] (char) {#1};}}
	\usepackage{array}
	\newcolumntype{C}[1]{>{\centering\let\newline\\\arraybackslash\hspace{0pt}}m{#1}}
	\newcolumntype{L}[1]{>{\let\newline\\\arraybackslash\hspace{0pt}}m{#1}}
	
	\newcommand{\tparagraph}[1]{\vspace{0.5\baselineskip}\noindent\textbf{{#1:}}}
	\usepackage{amsfonts}
	\usepackage[labelfont=bf]{caption}
	
	\usepackage{url}
	\makeatletter
	\def\Url@twoslashes{\mathchar`\/\@ifnextchar/{\kern-.2em}{}}
	\g@addto@macro\UrlSpecials{\do\/{\Url@twoslashes}}
	\makeatother
	\makeatletter
	\g@addto@macro{\UrlBreaks}{\UrlOrds}
	\makeatother
	\usepackage{csquotes}
	\usepackage[labelfont={bf}]{subcaption}
	\usepackage{algorithm}
	\usepackage{algpseudocode}
	\usepackage{enumitem} 
	\usepackage{amsmath}
    \usepackage{booktabs,multirow}
    \newcommand{\ra}[1]{\renewcommand{\arraystretch}{#1}}
    \newcolumntype{K}[1]{>{\centering\arraybackslash}p{#1}}

	\DeclareMathOperator*{\argmax}{arg\,max}
	
	\newcommand\negs{\vspace*{0\baselineskip}}
	\newcommand\posSpace{\vspace*{0\baselineskip}}
	\newcommand\snegs{\vspace*{0ex}}
   
	\usepackage{cuted}
	\PassOptionsToPackage{hyphens}{url}\usepackage{hyperref}
	\hypersetup{hidelinks}
	
\renewcommand{\footnotesize}{\small}
	
\begin{document}

		
        \permission{\footnotesize{This is the authors' extended version of the paper published at:}}
		\conferenceinfo{CODASPY'17,}{March 22 - 24, 2017, Scottsdale, AZ, USA}
		\isbn{978-1-4503-4523-1/17/03} 
		\doi{http://dx.doi.org/10.1145/3029806.3029837}

		\title{``If You Can't Beat them, Join them'': A Usability Approach to Interdependent Privacy in Cloud Apps}
		%
		%
		%
		%
		%
		
		\numberofauthors{2} 
		%

		\author{
			%
			%
			\alignauthor
			Hamza Harkous\\
			\affaddr{EPFL, Switzerland}\\
			\email{hamza.harkous@epfl.ch}
			\alignauthor
			Karl Aberer\\
			\affaddr{EPFL, Switzerland}\\
			\email{karl.aberer@epfl.ch}
		}

		\maketitle

		\begin{abstract}
			Cloud storage services, like Dropbox and Google Drive, have growing ecosystems of 3rd party apps that are designed to work with users' cloud files. 
			Such apps often request full access to users' files, including files shared with collaborators.
			Hence, whenever a user grants access to a new vendor, she is inflicting a privacy loss on herself and on her collaborators too. 
			Based on analyzing a real dataset of 183 Google Drive users and 131 third party apps, we discover that collaborators inflict a privacy loss which is at least 39\% higher than what users themselves cause.
			We take a step toward minimizing this loss by introducing the concept of \textit{History-based decisions}. Simply put, users are informed at decision time about the vendors which have been previously granted access to their data. 
			Thus, they can reduce their privacy loss by not installing apps from new vendors whenever possible.
			Next, we realize this concept by introducing a new privacy indicator, which can be integrated within the cloud apps' authorization interface. 
			Via a web experiment with 141 participants recruited from CrowdFlower, we show that our privacy indicator can significantly increase the user's likelihood of choosing the app that minimizes her privacy loss. 
			Finally, we explore the network effect of History-based decisions via a simulation on top of large collaboration networks. We demonstrate that adopting such a decision-making process is capable of reducing the growth of users' privacy loss by 70\% in a Google Drive-based network and by 40\% in an author collaboration network. This is despite the fact that we neither assume that users cooperate nor that they exhibit altruistic behavior.
			To our knowledge, our work is the first to provide quantifiable evidence of the privacy risk that collaborators pose in cloud apps. We are also the first to mitigate this problem via a usable privacy approach.

		\end{abstract}
		

		\section{Introduction}

		\tparagraph{The Rise of Cloud Apps}\\
		The popularity of consumer cloud storage providers (CSPs) over the previous decade has been on a roll. Dropbox, Google Drive, and One Drive have each amassed hundreds of millions of users. 
		In order to further appeal to their users, the CSPs have been transitioning from being pure \textit{service providers} to becoming \textit{app ecosystems}. Hence, they now offer APIs for developers to import and process users' files stored in the cloud.
		Consider, for example, a web app called \href{https://pandadoc.com}{PandaDoc},
		which allows creating, editing, and signing documents online. When a user uses PandaDoc from her laptop browser, she can import files stored in her Google Drive instead of her hard drive. Such a pattern is increasingly more prevalent with the growing number of 3rd Party Cloud apps (or 3PC apps) that are tightly integrated with cloud storage services. Dropbox alone claims that hundreds of thousands of apps
		have been integrated with its platform. 
		Even in the enterprise setting, 3rd party cloud apps are on the rise. This is first because companies are officially adopting the likes of \textit{Dropbox Business}, \textit{OneDrive for Business}, and \textit{Google Drive for Work}. Second, it is due to employees utilizing their personal cloud accounts to share company's files (a.k.a Shadow IT). Various reports from cloud application security providers state that organizations use from 10 to 20 times more cloud apps than their IT department thinks~\cite{SkyhighNetworks, ElasticaCloudThreatLabs2016}.

		\tparagraph{Risks in 3rd Party Cloud Apps}\\
		However, in our previous work, we have shown that 76\% of the 3rd party Google Drive apps featured on Google Chrome Store request full access to users' Google Drive data~\cite{Harkous16}. Around 64\% of these apps are \textit{over-privileged}: they require more permissions than are needed for them to function. 
		Accordingly, users are now faced with a new kind of privacy adversary: the 3rd party app vendors. With every app authorization decision that users make, they are trusting a new vendor with their data and increasing the potential attack surface. 
		Elastica, the cloud application security provider, estimates that the average financial impact on a company as a result of a cloud-storage data breach is \$13.85M, including remediation costs~\cite{Labs2015}. In 2015, the data breach at Anthem, a US insurance company, has reportedly cost more than \$100M, with 80M unencrypted health records leaked. This was a result of an exfiltration exploit leveraging a popular public cloud storage application~\cite{ElasticaCloudThreatLabs2016}. Even on the personal level, the risk extends from breaches exposing financial information and health records to unnoticeable, continuous profiling based on stored files. 
		
		
		\tparagraph{Exposure through Collaboration}\\
		An additional intricacy is that when users grant access to a 3rd party cloud app, they are not only sharing their personal data but also others' data. This is because cloud storage providers are inherently collaborative platforms where users share and cooperate on common files. Hence, protecting these files is not solely in the hands of the user. Skyhigh Networks, another provider of cloud security software, reports that 37.2\% of documents (across 23 million users) are shared with at least one other user. In organizations, documents are shared, on average, with accounts from 849 external domains~\cite{SkyhighNetworks2015}. 
		Moreover, around 23\% of cloud documents were found by Elastica to be ``broadly shared'', which means that they are shared (a) among all employees, (b) with external partners and clients, or (c) with the public~\cite{ElasticaCloudThreatLabs2016}. Interestingly, 12\% of those documents contained compliance-related or confidential data. 
		This further highlights what has been termed as the \textit{interdependent~privacy~problem}~\cite{biczok2013interdependent}, where the decisions of friends can affect the user's privacy and vice-versa. This concept was initially proposed in the context of third-party social networking apps, such as Facebook. However, while 1.92\% of Facebook apps request friends' personal information, this is much more pronounced in 3rd party cloud apps, where all apps accessing one's files get access to the part which is shared too. Moreover, unlike Facebook apps, due to the collaborative nature of cloud apps, the CSPs do not provide an option for users to control whether their collaborators' apps can get access to data they own.

		\tparagraph{Research Questions}\\
		So far, the main approach to reducing the risk of 3PC apps has been focused on discovering over-privileged apps and deterring users from installing them~\cite{Harkous16}. Even then, a lot of users would still install such apps as they prioritize short-term utility over long-term risk aversion or due to the absence of alternatives. Furthermore, that approach relies on manually inspecting each app by experts and on applying a plethora of machine learning algorithms to visualize the various risks for users. These issues could present a hurdle towards a wide-scale deployment by CSPs.
		In this work, we address the wider problem of minimizing the risk of all 3PC apps, regardless of whether they are over-privileged or least-privileged. We are further driven by the rationale that users will inevitably continue to install apps to achieve various services. Hence, instead of stopping them, we aim to lead them to select apps from vendors in a way that minimizes their privacy risk. We achieve this by leading users to take what we term as \textit{History-based decisions}. Such decisions account for the vendors who previously obtained access to the user's data, whether directly (with her consent) or via her collaborators. Our strategy consists of introducing privacy indicators to the current permissions interfaces that help users minimize the number of vendors with access to their data. Our ``usable privacy'' approach is guided via a data-driven study and is evaluated via a data-driven simulation.

		In essence, we tackle the following research questions:
		\begin{itemize}[leftmargin=*]
			\item
			From a practical perspective, are the collaborators' decisions significant enough to be accounted for in users' app adoption decisions?
			
			\item
			Do users already account for entities with access to their data? If not, to what extent can the usage of privacy indicators lead to users taking History-based decisions? 
			
			\item
			How significant is the effect of adopting these privacy indicators in the case of large networks of users and teams?
		\end{itemize}

		\tparagraph{Contributions}
		Towards addressing these questions, we make the following contributions:
		
		\begin{itemize}[leftmargin=*]
			
			\item
			In Section~\ref{sec:dataset}, we analyze a real-world dataset of Google Drive users, and we show that the median privacy loss that collaborators cause by installing apps can be much higher than that inflicted by the user's own app adoption decisions (39\% higher with 5\% of shared files and 523\% higher with 60\% of shared files). To our knowledge, this is the first usage of a real-world dataset to give a concrete evaluation of interdependent privacy in any ecosystem.
			
			\item
			Driven by the significant impact of collaborators, we design new privacy indicators for helping users mitigate the privacy risk via History-based  decisions (cf. Section~\ref{sec:study}). We assess these indicators via a web experiment with 141 users. We show that they significantly increase the likelihood that users choose the option with minimal privacy loss, even if not all of these users are motivated by privacy per se.
			To the best of our knowledge, this is also the first work to investigate a usable privacy approach to mitigating the problem of interdependent privacy. The few studies on this problem have mainly approached it from a theoretical perspective, such as developing game-theoretic or economic models~\cite{biczok2013interdependent,pu2014economic} or from a behavioral perspective, such as studying the factors affecting real users' monetary valuation of others' privacy~\cite{PuG15,PuG16}.

			\item
			We explore the potential of History-based decisions by performing a simulation on two large user networks. We show that the network-effects of our approach result in curtailing the growth privacy loss by 70\% in a synthetic Google Drive-based collaboration network and by 40\% in a real author collaboration network. We also simulate the effect of such decisions in a teams' network. We demonstrate that teams can reduce the privacy loss by up to 45\% by solely accounting for team members' decisions (cf. Section~\ref{sec:simulation}).

		\end{itemize}

		\section{Models and Preliminaries}
		\label{sec:models}
		
		\subsection{System Model}
		\label{sec:system}
		
		There are four main entities that interact in the third-party cloud app system: 
		\begin{enumerate}[leftmargin=*]
			\item
			a \textit{user} $u$ who uses that app for achieving a certain service
			\item
			a \textit{cloud storage provider (CSP)} hosting the user's \textit{data}
			\item
			a \textit{data subject} to whom the files belong and whose privacy is being considered. We further define two levels of data subject granularity: 
			\begin{itemize}[leftmargin=*]
				\item
				\textit{individual-level granularity}: i.e., the user herself is interested in guarding her own data privacy,
				\item
				\textit{team-level granularity}: i.e., a group of users are interested in guarding the privacy of team-owned data (e.g., using an enterprise version of cloud storage services)
			\end{itemize}
			\item
			a \textit{vendor} $v$ that is responsible for programming and managing a {3rd Party Cloud app} (or shortly a {cloud app} or a {3PC app}). 
			These vendors register their apps with the CSPs. The apps themselves are hosted on any website the vendors choose (\ie not hosted by the CSP itself).
			
		\end{enumerate}
		
		Each user has access to a set $F_u$ of files stored at the CSP.   A subset of these files is owned exclusively by the data subject while the other subset is composed of files that are each shared with at least one other \textit{collaborator}. 
		We denote the set of all collaborators of user $u$ by $C(u)$. For simplicity reasons, we will assume throughout this work that the files of all data subjects, as well as the collaborators for each file, are all fixed from a reference step $t=0$. 
		Using the CSP's API, the vendor $v$ can get access, at step $t \in \mathbb{N}$, to the subject's data upon \textit{user authorization}, which consists of $u$ accepting a list of \textit{permissions}. We will alternatively refer to this as \textit{app installation}, and we will assume that exactly one app is installed at each step $t$. Permissions are named differently across various providers, but, in general, we can categorize them into three categories:
		
		\begin{itemize}[leftmargin=*]
			\item
			\textbf{per-file access}: where the user has to authorize the vendor for each file access individually. This is typically done via a file picker provided by the CSP itself.
			
			\item
			\textbf{full-access}: where the vendor gets access to all users' data. In the interface, this is worded, for instance, as ``View the files in your Google Drive'' or ``access to the files and folders in your Dropbox''.
			
			\item
			\textbf{per-type access}: where the vendor gets access to all files of a specific type. For example, Dropbox words it as ``access to images in your Dropbox''. Some platforms, like Google Drive, do not provide app developers with such fine-grained options.
			
		\end{itemize}

		The authorization can also give $v$ access to files shared with the collaborators of $u$. Similarly, collaborators of $u$ can install apps that expose files shared with $u$ to new vendors. We denote the set of files of $u$ accessible by vendor $v$ at step $t$ as $F_{u,v}(t)$\footnote{Although we do not consider file deletion in this work, we note that, in the worst case, the vendor can still have access to copies of files it saved before the user deleted them.}. 
		
		\begin{table}[t]
			\centering
			\begin{tabular}{ll}
\toprule

            \textbf{Notation}         & \textbf{Explanation}                          \\
            \midrule
				$u$         & user                                         \\
				$v$         & vendor                                       \\
				$C(u)$       & Collaborators of $u$                         \\
				$V_u$       & set of vendors authorized by $u$             \\
                				$V_{c(u)}$       & set of vendors authorized by collaborators of $u$     \\        
				$\VFC_u(V)$ & file coverage due to the vendors in set V    \\
				$F_{u}$     & set of files of $u$                          \\
				$F_{u,v}$   & set of files of $u$ accessible by vendor $v$\\
                             \bottomrule

			\end{tabular}
            \caption{Summary of notations used}
			\label{tab:notation}
		\end{table}
		
		\subsection{User Model}
		\label{sec:userModel}
		A user is further assumed to be  \textit{self-interested}, \ie only caring about optimizing the privacy of the data subject (a.k.a., privacy egoist), and \textit{non-cooperative}, \ie does not coordinate her decisions with others. 
		We do not assume that the risks of installing each app are known to the users or calculated a priori. 
		In fact, unlike other 3rd party app ecosystems, the risk of each cloud app cannot be automatically estimated based on techniques such as taint tracking~\cite{Enck:2010} or code analysis~\cite{Felt:2011:APD} because the main app's functionality is typically implemented on the server side (which cannot be accessed by external entities). 
		Such assumptions constitute the \textit{worst case} in the scenarios we consider, and further privacy optimizations can be obtained by relaxing them.
		
		We also assume that the mental model for privacy-concerned users matches the possible permission granularities they are given. Accordingly, privacy-concerned users can have one of the following privacy-goal granularities\footnote{Per-file access already achieves the least privilege possible.}:
		\begin{itemize}[leftmargin=*]
			\item
			\textbf{per-type privacy goal}: where users aim to optimize their privacy independently for different file types. For example, in an ecosystem like Dropbox, where per-type access is an option, users might follow the separation-of-concerns principle. Hence, they might install photo-related apps from a set of vendors that is different from the set authorized for document processing.
			\item
			\textbf{all-files privacy goal}:  where users aim to reduce the privacy risk for their entire set of files. This can be in the case of ecosystems which do not have the option of per-type access, like Google Drive. It can be also the case that a user of Dropbox has this goal in mind despite being presented with finer-grained app permissions. 
		\end{itemize}
		
		\subsection{Threat Model}
		We consider the 3rd party app vendors as the adversary (and not the CSP). The privacy indicator we introduce is best implemented by the CSP, which already has access to the users' and collaborators data.  Alternatively, this can be a feature within Cloud Access Security Brokers (e.g., SkyHigh Networks, Netskope, etc.), which are already trusted by thousands of enterprises to protect their cloud data against other 3rd parties. Moreover, we consider the protection against over-privileged apps as an orthogonal problem, which we have considered in~\cite{Harkous16}. We rather focus on the interdependent privacy problem, which covers all vendors with full access and is an issue in least-privileged apps too.
		
		\subsection{Privacy Loss Metrics}
		\label{sec:privloss}
		In order to quantify the privacy loss that a user incurs with time, we introduce now the \textit{Vendors File Coverage ($\VFC$)} metric. Consider a user $u$ and a set $V$ of vendors at a certain time step. For notation simplicity, we will omit the time step henceforth. $\VFC_u(V)$ is computed as the summation of the files' fractions shared with each of these vendors:
		\snegs
		\begin{equation}\label{equation:VFC}
		\VFC_u(V) = \sum_{v\in V} \frac{|F_{u,v}|}{|F_u|}
		\end{equation}

		Intuitively, $\VFC_u(V)$ increases as vendors in $V$ get access to more files of $u$. It has the range $[0,|V|].$~\footnote{We do not normalize $\VFC_u(V)$ by $|V|$ as multiple vendors with access to all the user's files induce a higher privacy loss than one vendor with such access.}
		
		If we consider the set $V_u$ of vendors explicitly authorized by user $u$, we can define the \textit{Self-Vendors File Coverage} as:
		\begin{equation}
		\mbox{\textit{Self-}}\VFC_u= \VFC_u(V_u)
		\end{equation}

		Similarly, if we consider the set $V_{C(u)}$ of vendors  authorized by the collaborators $C(u)$ of $u$, we can define the \textit{Collaborators-Vendors File Coverage} as:
				\begin{equation}
\mbox{\textit{Collaborators-}} {\VFC_u=\VFC_u(V_{C(u)})}
				\end{equation}
		Finally, the \textit{Aggregate $\VFC_u$} for a user $u$ is that due to all vendors authorized by $u$ or its collaborators:
				\begin{equation}
			\mbox{\textit{Aggregate-}}	\VFC_u = \VFC_u(V_u \cup V_{C(u)})
				\end{equation}\label{equation:AggVFC}

		Throughout this work, we will use the terms \textit{privacy loss} and $\VFC$ interchangeably. As will become evident in Section~\ref{sec:study}, this metric choice allows relaying a message that is simple enough for users to grasp, yet powerful enough to capture a significant part of the privacy loss. Obviously, one can resort to a deeper inspection of content or metadata sensitivity (as in~\cite{harkous2014c3p}) had the purpose been finding the best privacy model in general. However, for instigating a behavioral change, telling users that a company has 30\% of their files is more concrete than a black-box description informing them that the calculated loss is 30\% and constitutes less information-overload than presenting them with detailed loss metrics.
		
		\posSpace

		\section{Collaborators' Impact}
		\label{sec:dataset}
		At this point, we are in a position to handle the first research question on the extent of collaborators' contribution to a user's privacy loss. Hence, we want to test the following hypothesis:
		
		\textit{H1: The collaborators' app adoption decisions have a significant impact on the user's privacy loss.}
		
		If this hypothesis is valid in practice, it provides a strong motivation for designing privacy notices that aid users in accounting for their collaborators' decisions, which is what we will study in Section~\ref{sec:study}.
		Towards that, we will be dissecting the privacy loss, quantified by $\VFC$, that users incur in a realistic 3rd party cloud apps dataset.

		\subsection{The Case of Google Drive}

\begin{figure}[t]
			\centering
			\fbox{\includegraphics[width=0.9\linewidth]{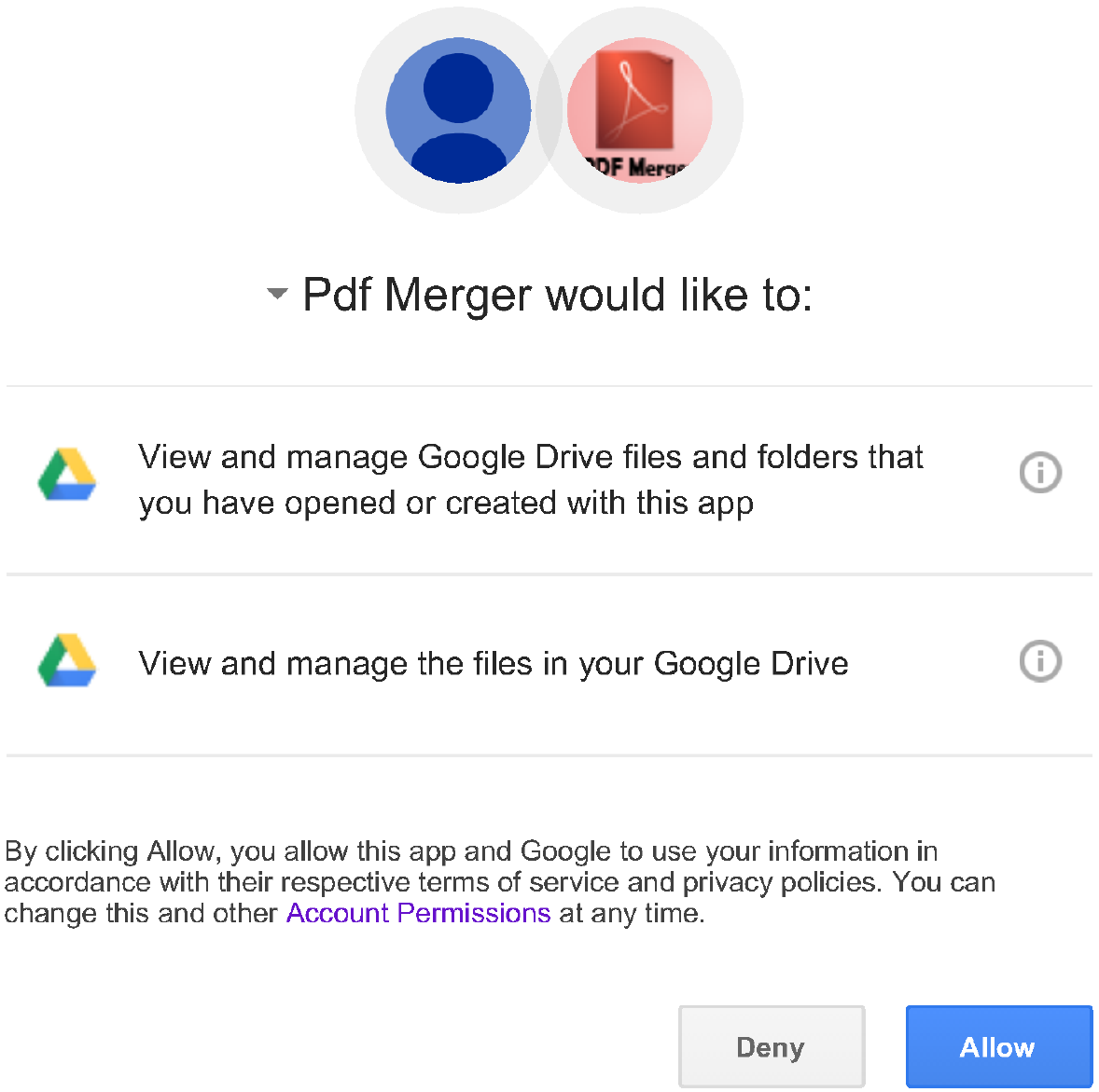}}
			\caption{Current permissions interface of Google Drive}
			\label{fig:baseline}
		\end{figure}

		To study the problem in a realistic context, we will be taking Google Drive as a case study in this work, given that it has one of the most popular 3rd party ecosystems. Nevertheless, the insights gained from our work are applicable to other cloud platforms as well. 
		The main (content-related) Google Drive permissions that 3PC apps' vendors can request are presented Table~\ref{tab:defaultPermissions}, along with the Google-provided description for each. This short description is also presented to the user when installing an app (see Figure~\ref{fig:baseline} for an example app). 
		The user can click on the info button \circled{i} next to each permission to read additional explanations in a popup. 
		The user has to accept all permissions in order to utilize the app. These apps can be found on Google Chrome Web Store (and other Google stores), where users can rate and review them. 
		In this work, we will focus on content-related permissions. Hence, as discussed in Section~\ref{sec:models}, we differentiate between two levels of access: (1) full access, which includes the \textsc{drive\_readonly} and \textsc{drive} permissions and (2) per-file access that includes the \textsc{drive\_file} permission. Google Drive does not offer the per-type permissions option.
		\begin{table}[t]
        \small
			\centering
			\begin{tabular}{L{4.5cm}cl} 
				\toprule
                \textbf{Permission} & & \textbf{Short Name}\\
                \midrule
                \\
                 \sffamily View the \textbf{files} in your Google Drive.& & \textsc{drive\_readonly}\\ 
                \\
				\sffamily View and manage the \textbf{files} in your Google Drive.& & \textsc{drive}\\ 
                \\
			
				\sffamily View and manage Google Drive files that you have \textbf{opened or created with this app}. & &  \textsc{drive\_file}\\ 
				\\
                \sffamily View your Google Drive {apps}. & & 
				\textsc{drive\_apps\_readonly}
                \\ 
			
				\bottomrule
			\end{tabular}
			\negs
			\captionof{table}{Requested permissions with the short reference name}
			\label{tab:defaultPermissions}
			\snegs
		\end{table}

		\negs
		\subsection{Dataset}
		\label{sec:privsealDataset}
		
		One of the main challenges when studying the privacy loss in 3rd party cloud apps is the absence of public datasets with realistic file distributions, collaborator distributions, sharing patterns, 3rd party app installations, etc.
		We benefit in this section from a dataset that we have collected in a previous work via the PrivySeal\footnote{\url{https://privyseal.epfl.ch}} service~\cite{Harkous16}. We build our analysis on it in order to evaluate the $\VFC$ of users in a realistic context. 
		The dataset, henceforth referred to as the \textit{PrivySeal Dataset}, was anonymized and contained metadata-only information.
		It included a subset of the files' metadata of 183 PrivySeal users in addition to the Google Drive apps installed by those users prior to authorizing PrivySeal's app (the \textsc{drive\_apps\_readonly} permission was requested by PrivySeal).
		Each user had a minimum of $N_{files\_min}=10$ files in total and at least $P_{min\_shared}=5\%$ of files that are shared. The dataset specifically contained:
		
		\begin{itemize}[leftmargin=*]
			
			\item 
			list of user IDs (anonymized via a one-way hash function);
			\item
			IDs of files in each user's Google Drive,
			\item 
			list of anonymized collaborators' IDs for each file ID;
			\item
			list of apps with full-access installed by each user;
			\item 
			the vendor of each app.
		\end{itemize}
		
		In total, the number of users in addition to collaborators was 3422. Overall, these users had installed 131 distinct Google Drive apps from 99 distinct vendors. 
		Figure~\ref{fig:datasetStats} characterizes the PrivySeal Dataset. Particularly, it displays 4 distributions in this dataset, which realistically model the system under study:

		\begin{itemize}[leftmargin=*]
			\item
			number of files per user, which follows a skewed distribution with a median of 67 files
			\item
			sharing pattern: percentage of shared files out of all user files, which also follows a skewed distribution with a median around 18\%
			\item
			number of collaborators across all user files (a.k.a., the degree of the user node in the collaboration network): where 75\% of the users had less than 23 collaborators
			\item
			number of vendors authorized per user: also follows a skewed distribution with a median of 1 vendor per user
		\end{itemize}

		\subsection{Results}
		\label{sec:dataResults}
		
\begin{figure}[t]
			\centering
			\begin{minipage}{.45\textwidth}
				\begin{subfigure}[b]{0.5\linewidth}
					\centering
					\includegraphics[width=0.9\linewidth]{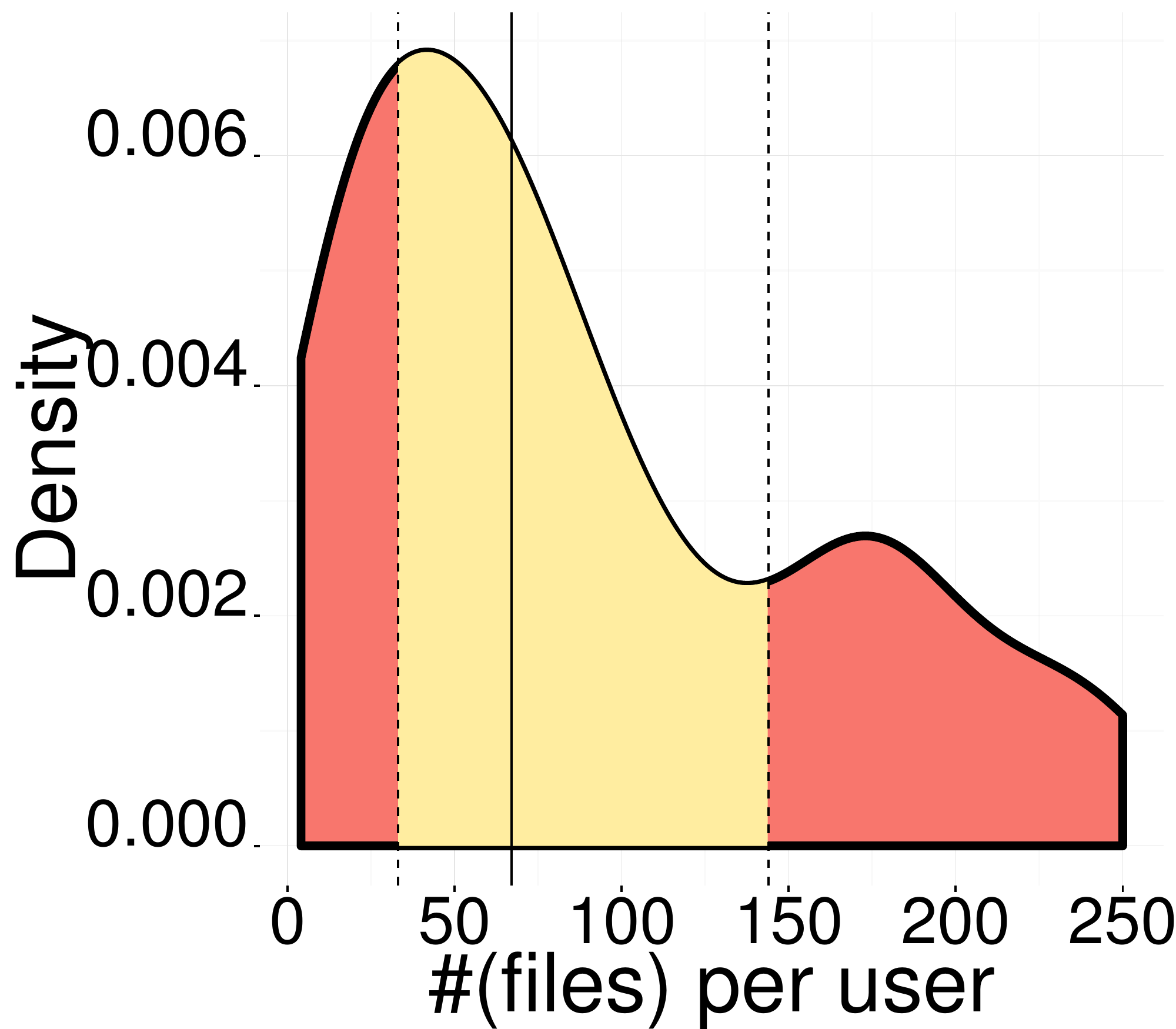} 
					
					\caption{Files' count} 	
					\label{fig:dataset-fileCountDistro} 
					
				\end{subfigure}
				\begin{subfigure}[b]{0.5\linewidth}
					\centering
					\includegraphics[width=0.9\linewidth]{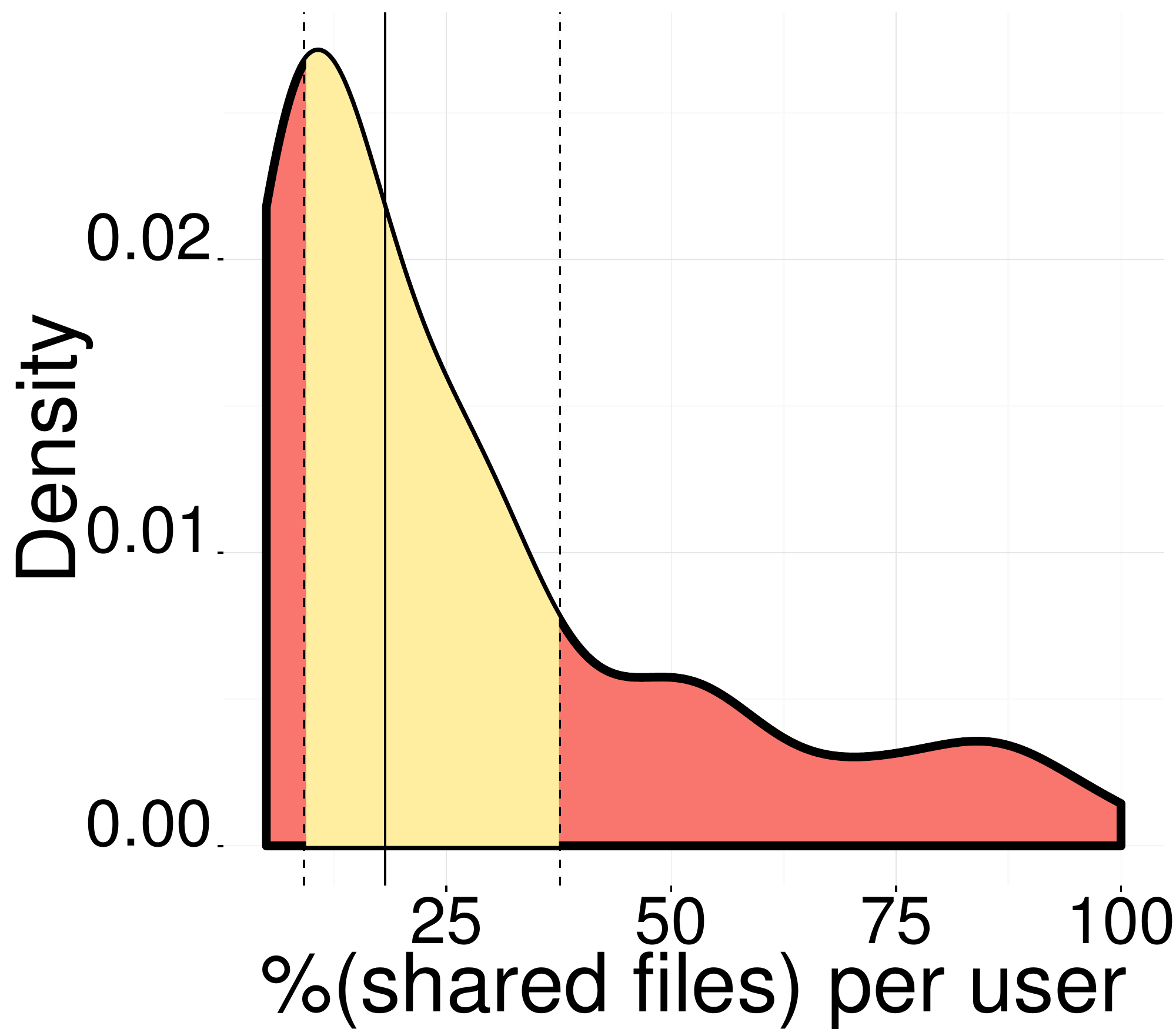} 
					
					\caption{shared files' \% } 
					\label{fig:dataset-fileSharingDistr} 
					
				\end{subfigure} 
				\begin{subfigure}[b]{0.5\linewidth}
					\centering
					\includegraphics[width=0.9\linewidth]{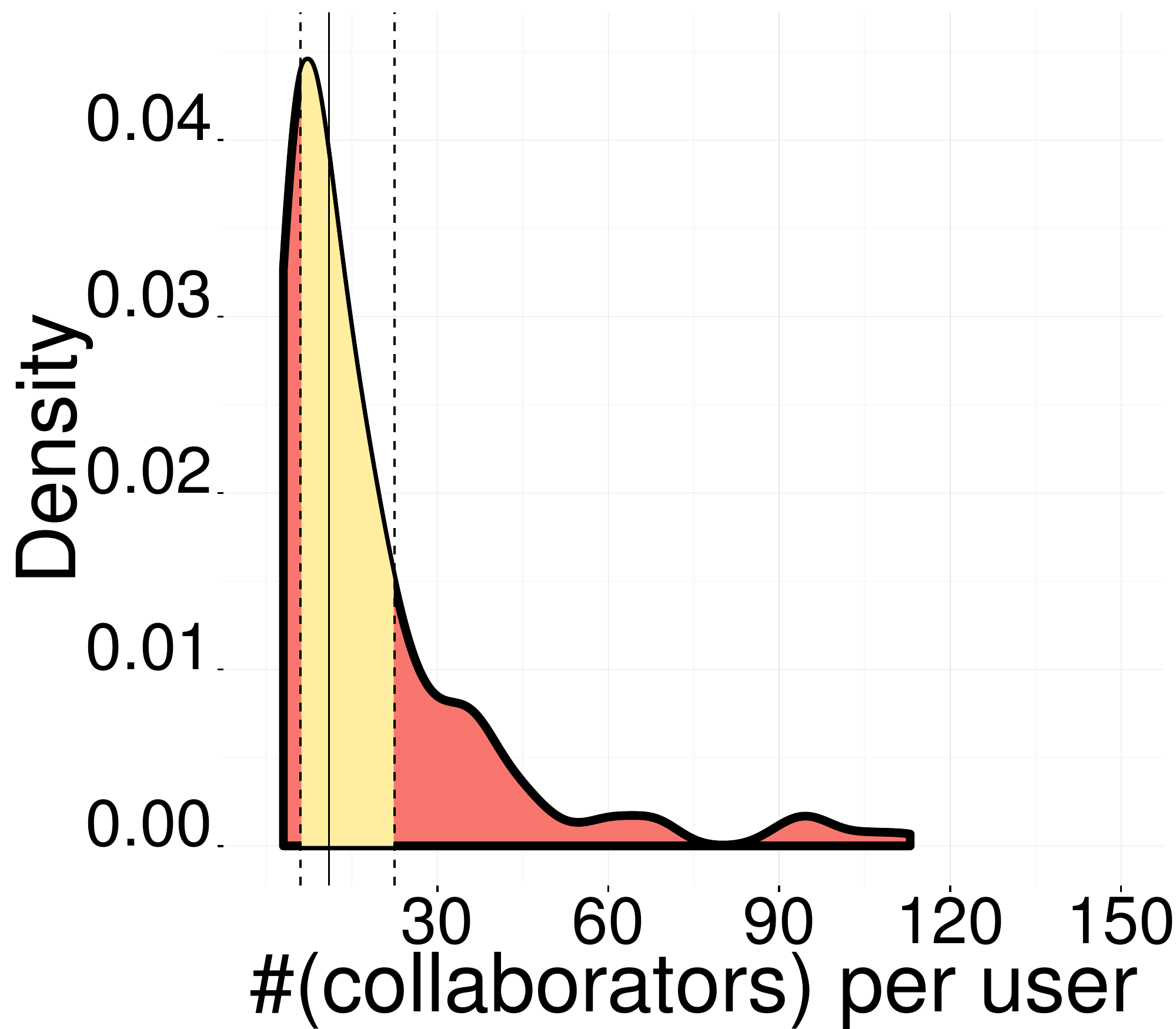} 
					\caption{Collaborators' count} 
					\label{fig:dataset-collabCount} 
				\end{subfigure}
				\begin{subfigure}[b]{0.45\linewidth}
					\centering
					\includegraphics[width=0.9\linewidth]{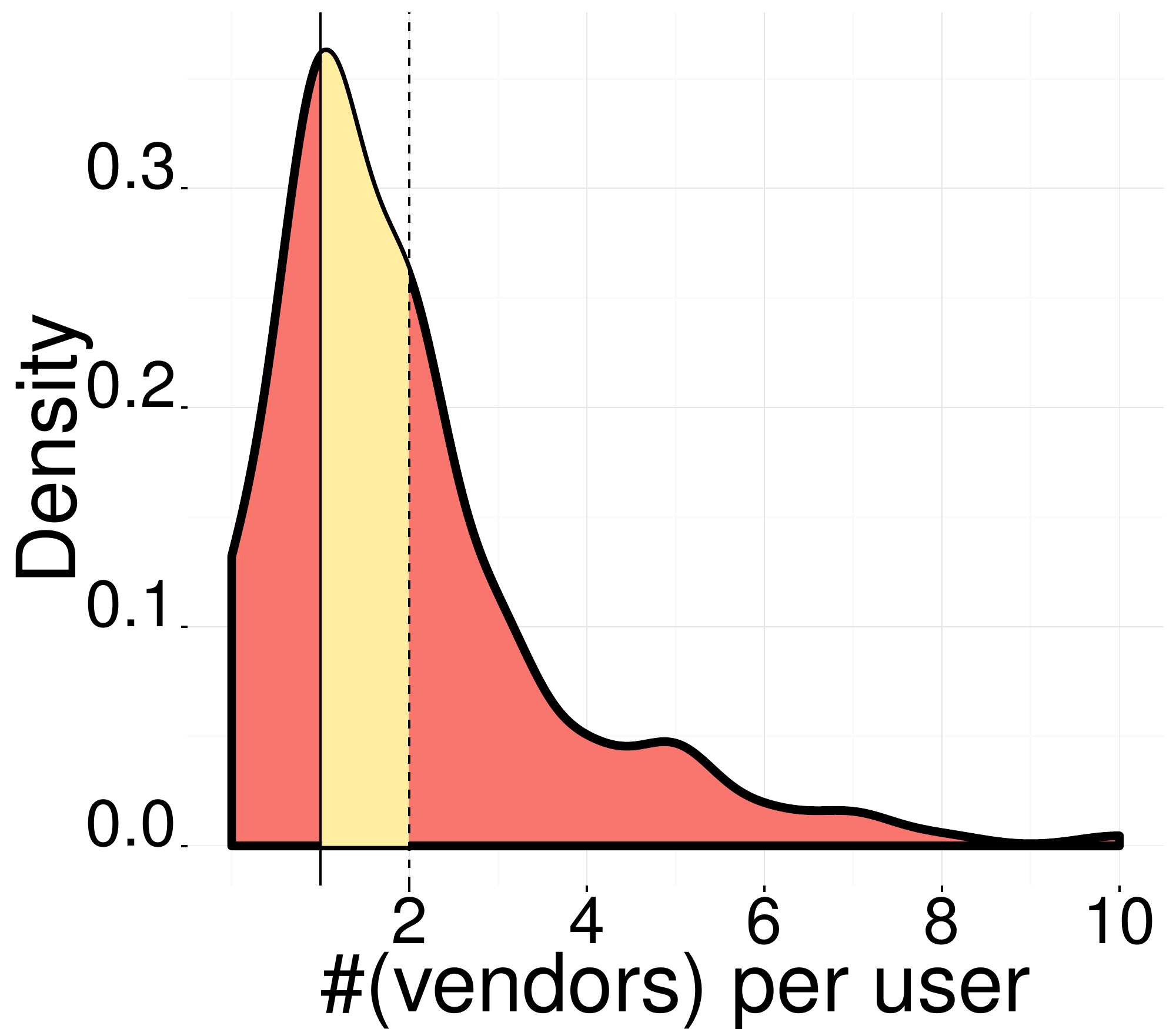} 
					\caption{Vendors' count} 
					\label{fig:dataset-fullVendorsCount} 
				\end{subfigure} 
				\caption{Density plots for various parameters, computed per user ($P_{shared\_min}=5$). Median line is shown, and the light orange area represents the range between the 25\% to 75\% quantiles. }
				\posSpace
				\label{fig:datasetStats} 
			\end{minipage}
			\hspace{1ex}
			\begin{minipage}{0.5\textwidth}
				\centering	
				\includegraphics[width=0.8\textwidth]{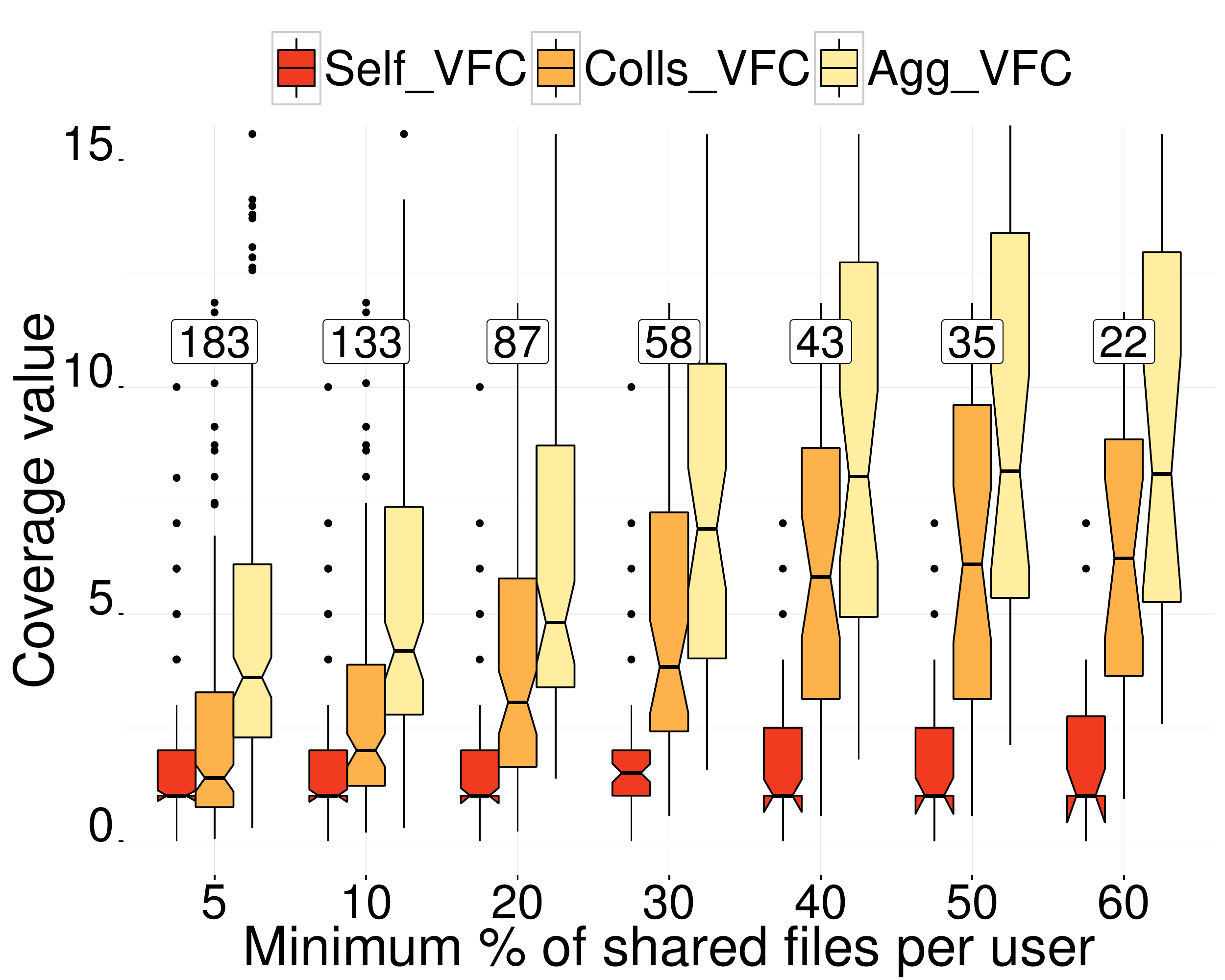}
				\caption{Evolution of metrics with populations that share more files ($N_{files\_min}=10$, $N_{apps\_min}=1$). The numeric labels denote the corresponding number of users in the dataset.}
				\label{fig:coverage_growth_withSharing}
			\end{minipage} 
		\end{figure}

		We computed the \textit{Self-$\VFC$}, the \textit{Collaborators-$\VFC$}, and the \textit{Aggregate-$\VFC$} (as defined in Section~\ref{sec:privloss}) for users in the PrivySeal Dataset\footnote{To avoid double counting, we considered the vendors authorized by both the user and her collaborators in computing \textit{Self-$\VFC$} but not in computing \textit{Collaborators-$\VFC$}.}. 
		As we did not have the actual number of apps for each collaborator of users in the dataset, we assigned to these collaborators a set of apps from a random user of the dataset.
		We show in Figure~\ref{fig:coverage_growth_withSharing} how these metrics evolve as we gradually consider populations that collaborate more frequently. With $P_{min\_shared}=5\%$, we had a median of 1.39 for  \textit{Collaborators-$\VFC$}, which was 39\% higher than a median of 1.00 for \textit{Self-$\VFC$}. The significance of the median difference is evidenced by the non-overlapping box-plot notches. This difference became much larger when we considered users that share more files. We had a 100\% median difference at $P_{min\_shared}=10\%$ and 523\% median difference at $P_{min\_shared}=60\%$. 
		Such results indicate that:
		\begin{itemize}[leftmargin=*]
			\item
			The {collaborators' app adoption decisions contribute a core component to the user's privacy loss}, thus confirming our hypothesis $H1$.
			\item
			{The higher the number of collaborators is, the higher the magnitude of loss these collaborators} can potentially inflict.
		\end{itemize}
		Both conclusions motivate the need for taking collaborators' decisions into account when designing privacy indicators for cloud apps, which is what we will embark on next.

		\section{User Study}
		\label{sec:study}
		
		Up till now, we have confirmed that, if users want to minimize their privacy loss, they are better off not ignoring the app installation decisions of collaborators. 
		In this section, we tackle the next research question, where we investigate the potential of privacy indicators in leading users to minimize their exposure to 3PC app vendors. 
		We show first our design methodology for the privacy indicators, and we follow that by a web experiment that investigates the efficacy of these indicators in realistic scenarios.
		
		\subsection{History-based Privacy Indicators}
		\label{sec:HBdesign}

\begin{figure}[t]
			\centering	
			\fbox{\includegraphics[width=0.9\linewidth]{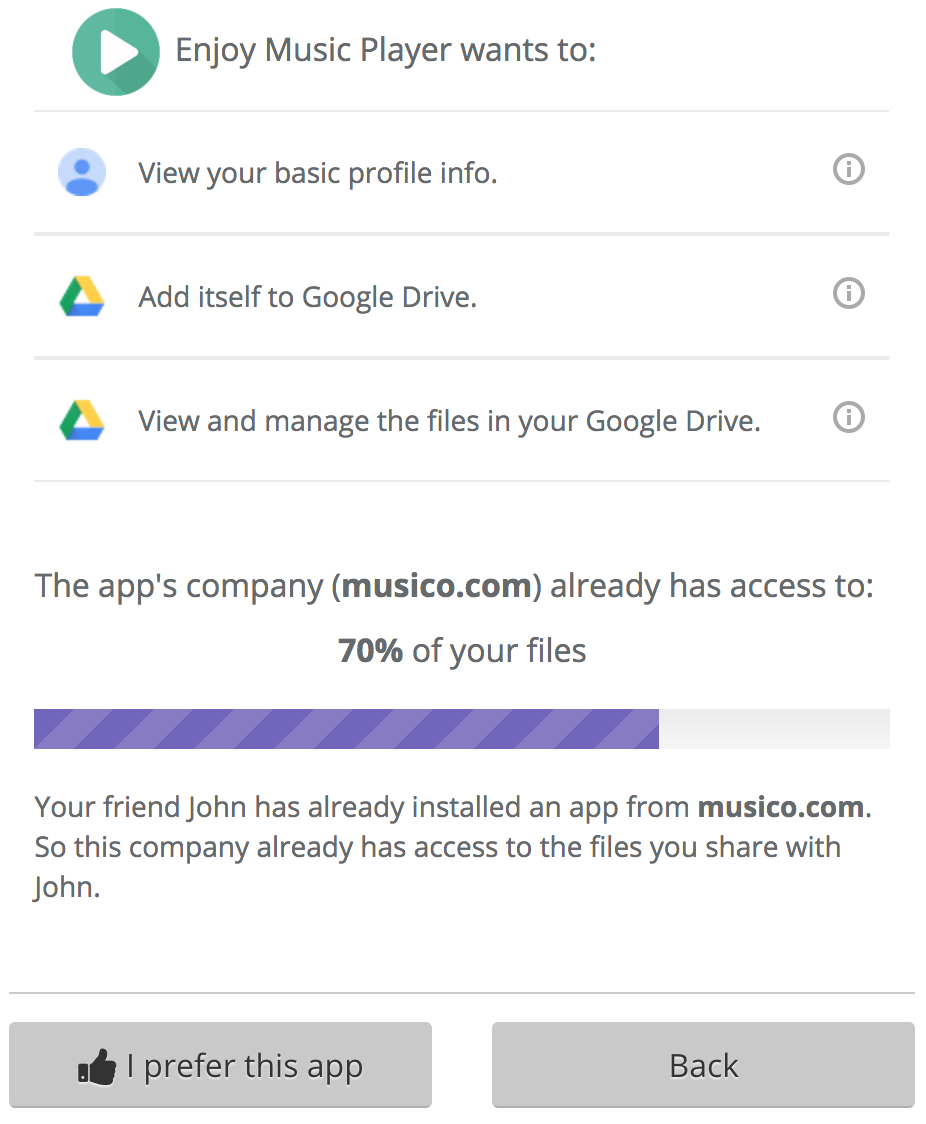}}
			\caption{Proposed ``History-based insights'' interface, with the buttons from the user study in the bottom}
			\label{fig:history_interface}
			\negs
		\end{figure}

		We call our proposed privacy indicators \textit{``History-based Insights'' (\HB\ Insights)} as they allow users to account for the previous decisions taken by them or by their collaborators. 
		We continue to consider Google Drive as a case study, and we show this indicator in the context of Google Drive apps' permissions in Figure~\ref{fig:history_interface}. Compared to the current interface provided by Google (Figure~\ref{fig:baseline}), we added a new part to highlight the percentage of user files readily accessible by the vendor (computed based on $\VFC_u(\{ v \})$ for each vendor $v$).
		As we prove in Appendix~\ref{sec:optimal}, selecting the vendor that already has the largest percentage of user files is the optimal strategy to minimize the privacy loss in our context. We denote this strategy as \textit{``History-based decisions''}.
		Following the best practices in privacy indicators' design~\cite{Design_Schaub}, our indicator was multilayered, with both textual and visual components. The wording of the main textual part was brief and general enough to hold for both the data percentage exposed by friends and that exposed by the user. We used a percentage value rather than a qualitative measure to facilitate making comparisons among apps based on this value. The visual part showed the percentage as a progress bar with a neutral violet color. The bottom textual part was added in a smaller font to provide further explanation for those interested. We used the term \enquote{company} in our interface instead of \enquote{vendor} as it is more commonly understood by the general audience.

		\begin{figure}[t]
			\centering	
			\fbox{\includegraphics[width=0.8\linewidth]{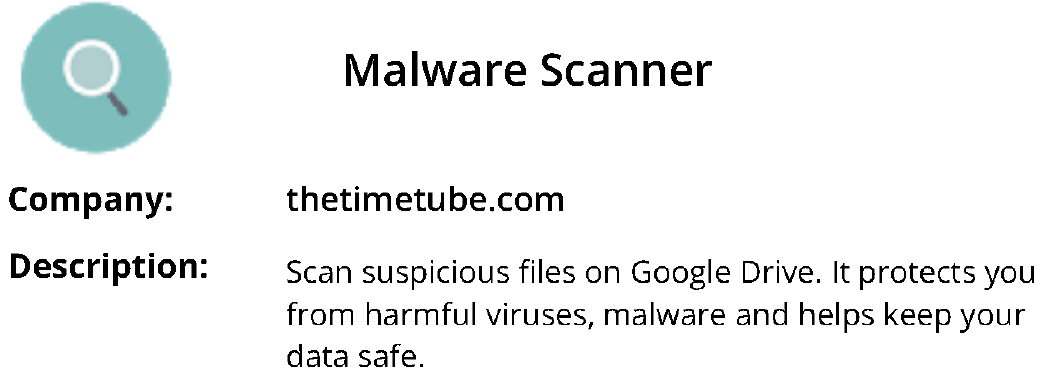}}
			\snegs
			\caption{Example app displayed in the list of apps}
			\label{fig:app_info}
		\end{figure}

		\subsection{Methodology}
		In order to evaluate the new permissions interface, we performed an online web experiment (rather than a lab study) as we were mainly motivated by obtaining a large sample of users that is also geographically and culturally diverse. 
		The hypothesis we wanted to test is:
		
		\textit{H2: Introducing the new privacy indicator significantly increases the probability that users take History-based decisions. }
		
		In addition, the study allowed us to build a realistic user decision model based on the choices taken by participants in different conditions. We will utilize this model in Section~\ref{sec:simulation} to simulate the app choices in a large user network and to study the effect on the overall $\VFC$ in the network. 
		We structured our study to have (1) an Introductory Survey, (2) a series  of App Installation Tasks, and (3) a Concluding Survey.

		\tparagraph{User Recruitment}
		We recruited users via CrowdFlower's crowdsourcing platform. In our study, we restricted participation, via the platform's filtering system, to the highest quality contributors (Performance Level 3). We also geographically targeted countries where English is a main language as our interface was only in English. In order to further guarantee quality responses, each user was rewarded a small amount of $\$ 0.5$ for merely completing the study and an additional amount of $\$ 1.25$ that was manually bonused for those who did not enter irrelevant text in the free-text fields.
		
		\tparagraph{Instructions}
		Participants were first presented with introductory instructions that explained the context of the study (\ie cloud storage services and 3rd party apps that can be connected to them). They were asked to only continue if they had good familiarity with cloud storage services (\eg Google Drive, Dropbox, etc.). We did not explicitly require that participants have experience with 3rd party cloud apps. However, we educated them about such apps throughout the instructions, particularly showing them two examples of 3rd party apps in action (PandaDoc for signing documents and iLoveIMG for cropping photos). These apps were displayed via animated GIFs that play automatically and do not rely on the user clicking.
		We used limited deception by neither mentioning the focus of the study on participants' privacy nor giving hints about selecting apps based on the installation history. The advertised purpose was to \enquote{check how people make decisions when they install 3rd party apps.} 
		
		\tparagraph{Introductory Survey}
		After checking the instructions, users were presented with an introductory survey, where they first entered general demographic information. This survey was also front-loaded with questions about cloud storage services (several of which required free-text input) in order to discourage users who had not used these services from continuing to the actual study.

		\subsection{Study Overview}
		Next, users could proceed to the study page. We used a split-plot design in the study. Participants were randomly assigned to one of two groups: 
		
		\begin{enumerate}[leftmargin=*]
			\item
			\textbf{Baseline Group (\BL)}: where the permissions interface used is that currently provided by Google Drive (Figure~\ref{fig:baseline}).
			\item
			\textbf{History-based Group (\HB)}: where the \hbi permissions interface (Figure~\ref{fig:history_interface}) is used.
		\end{enumerate}
		
		In each group, the study consisted of 3 modules, which cover the main conditions that can occur when users desire to install a cloud app.
		On a high level, the modules investigate the following questions:
		\begin{enumerate}
			\item
			\textbf{Module 1:} are users likely to select apps from the same vendor they installed from before?
			\item 
		\textbf{Module 2:} are users likely to select apps from vendors that her collaborators have used before?
			\item
		\textbf{Module 3:} do users consider the differences in access levels obtained by vendors that collaborators installed?
		\end{enumerate}
		
		In all modules, whenever the user was asked to \textit{choose} an app, she was presented with a list of 12 apps (Figure~\ref{fig:app_info} shows an example app). Only two of these apps were relevant to the task purpose, and they were placed on top of the list (randomly positioned as first or second). With this setup, we wanted to mimic the realistic setup of app browsing while not squandering the user's effort on finding apps. All apps had the same full access permissions too (namely \textsc{drive} permission). Unlike in Chrome Store, we removed elements such as ratings, user reviews, and screenshots and kept a minimal interface. This is all in order to reduce the distractions from factors outside the study. We refer the user to the work of Kelly et al.,~\cite{Kelley:2013} who investigated the effects of those elements on users' decisions for Android apps.
		
		In order to account for fatigue and learning effects, modules 1, 2, and 3 were presented in a random order for users. 
		We piloted our experimental setup in two stages: with colleagues and with online users from the CrowdFlower community itself. For reviewing the online pilot testers work, we embedded a Javascript code for session recording in our study's web page, which allowed us to view the user's mouse and keyboard actions on our side. 
		
		\tparagraph{Demographics} We had 157 users who completed the study. Based on manually reviewing the users' inputs, we removed 16 users who were inputting irrelevant free-text in the survey in the study. We thus report the results of 141 users, 72 of which were in the \BL\ group and 69 in the \HB\ group. In~Table~\ref{tab:demographics}, we describe the participants' demographics based on the introductory survey. 
		Of these participants, $66.4\%$ were males and $33.6\%$ were females. They were between 18 and 62 years old, with a median of 31. Moreover, 42.3\% of the participants had worked or studied in IT before. Participants were mostly from India (37\%), USA (35\%), Britain (7\%), Germany (7\%), and Canada (7\%).
		
		CrowdFlower presents the users with an optional satisfaction survey after completing the study, and 49 users took this survey.  On average, the study received 4.2/5 for instructions clarity, 3.8/5 for questions' fairness, 3.8/5 for ease of job, 3.6/5 for pay sufficiency (before the bonus was rewarded). This ensures that participants' behavior has not been affected by either a lack of time to complete the task or the task design in general.
		
        \begin{table}[t]
			\footnotesize
			\begin{tabular}{r  l  l  }
            \hline\\
				\textbf{Age} & 18-62  & 	(median 31 years)  \\ 
				& & \   \\ 
				\textbf{Gender} & 35.5\% & Female \\ 
				&	64.5\% & Male   \\ \\
                
				\textbf{Occupation} & 59.6\% & full-time employees \\ 
				&	14.2\% & student   \\ 
				&	6.4\% & part-time worker   \\ 
				&	8.5\% & self-employed   \\ 
				&	5.0\% & homemaker  \\ 
				&	6.4\% & Unemployed/retired   \\ \\
                
				\textbf{IT Experience} & 41.8\% & Have worked or studied in IT  \\ \\
				\textbf{Degree} & 19.1\% & High school \\ 
				&	7.1\% & Trade/tech./vocational training   \\ 
				&	51.1\% & Associate or Bachelor's degree   \\ 
				&	22.7\% & Post Graduate Degree   \\ \\
				\textbf{Countries} & 35.0\% & USA \\ 
					& 37.5\% & IND   \\ 
					&7.5\% & GBR   \\ 
				&	6.9\% & DEU   \\ 
				&	6.9\% & CAN   \\ 
				&	7.4\% & AUS+IRL+ NLD + PAK   \\  
				\hline
			\end{tabular}
			\caption{Demographics in our user study; $N=141$}
			\label{tab:demographics}
		\end{table}

		\subsection{Study Details and Results}

\begin{figure}[t]
\centering
\framebox{\includegraphics[width=1\linewidth]{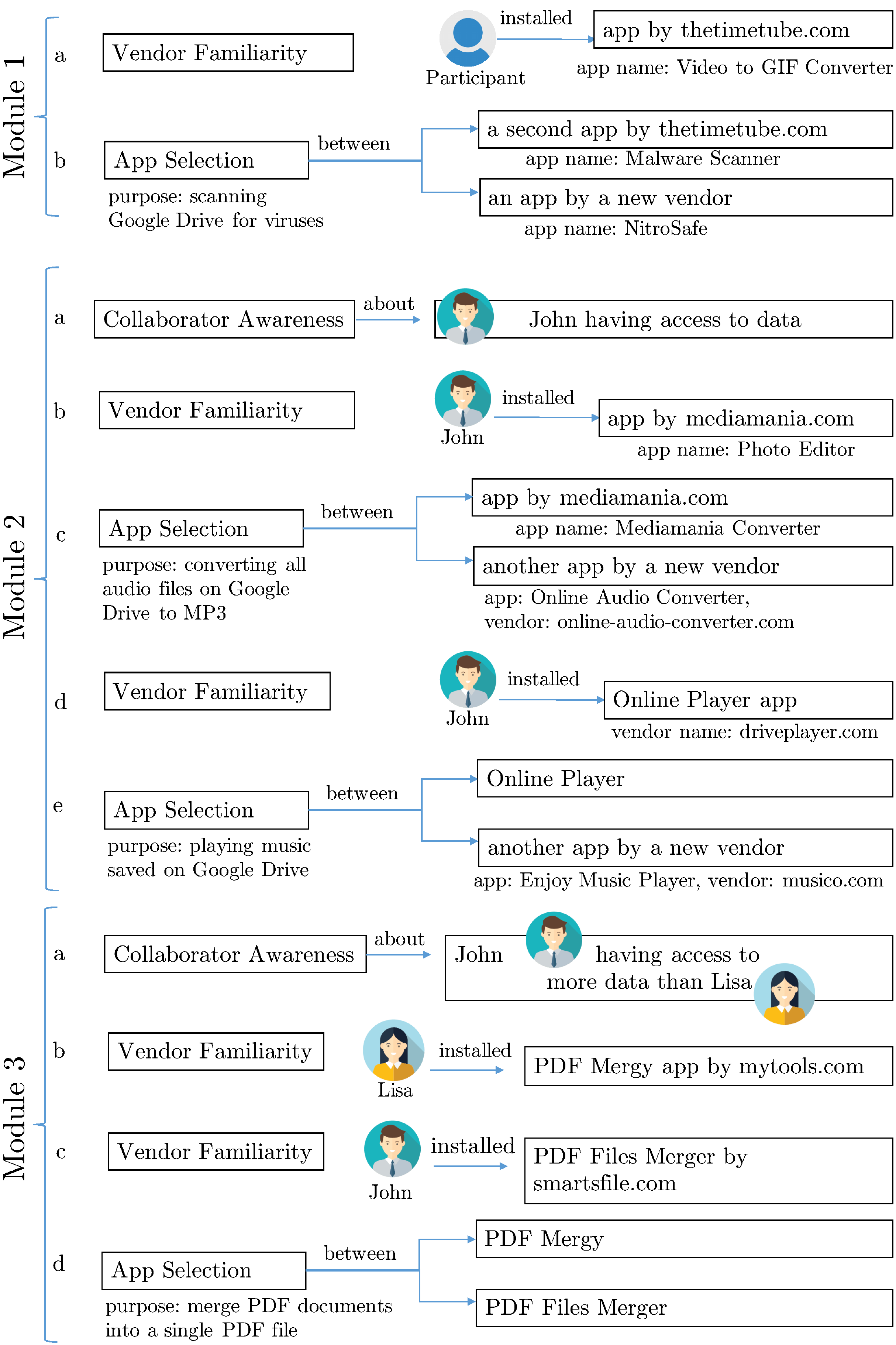}}
\caption{Summary of the experiment modules; a sample of the questions corresponding to each step are available in Figure~\ref{fig:screenshots}. 
}
\label{fig:modules}
\end{figure}

\begin{figure}[t]
{	
\minipage{1\linewidth} 
\centering
\framebox{\includegraphics[width=0.96\linewidth]{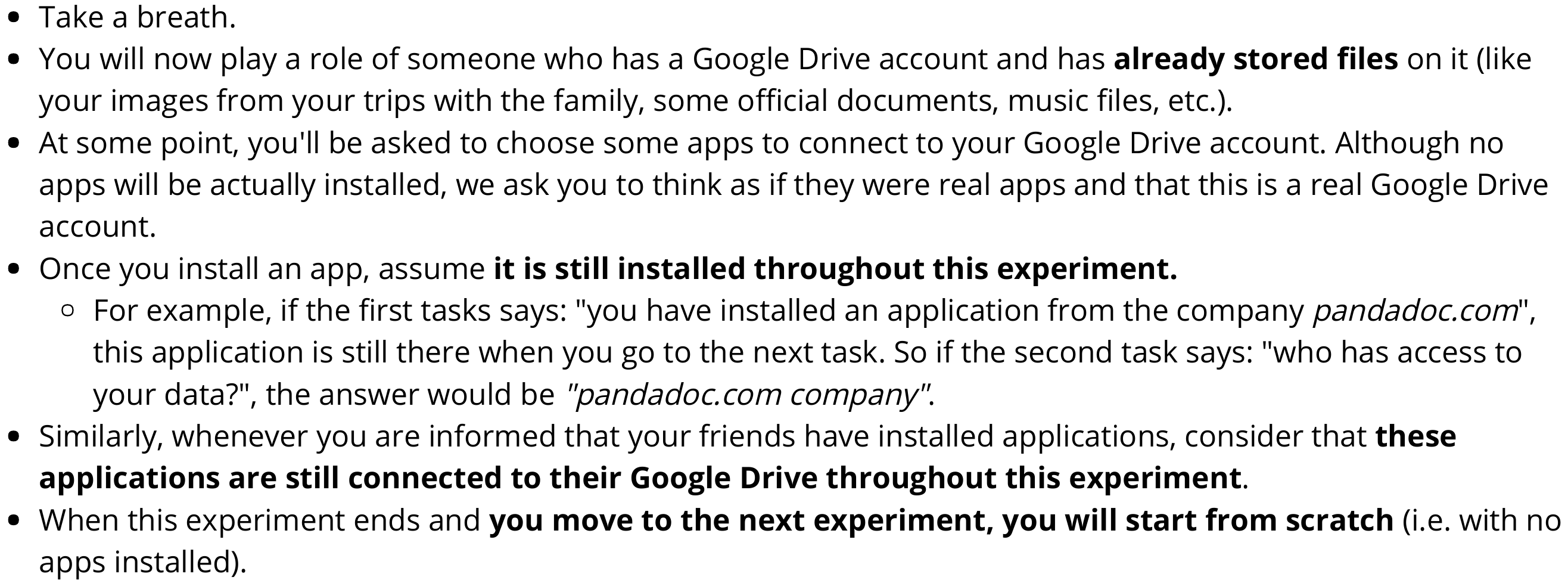}}
\subcaption{Instructions presented to users at the beginning of a module}
\label{fig:instructions}
\vspace{0.7\baselineskip}
\endminipage\\
\minipage{1\linewidth} 
\centering
\framebox{\includegraphics[width=0.96\linewidth]{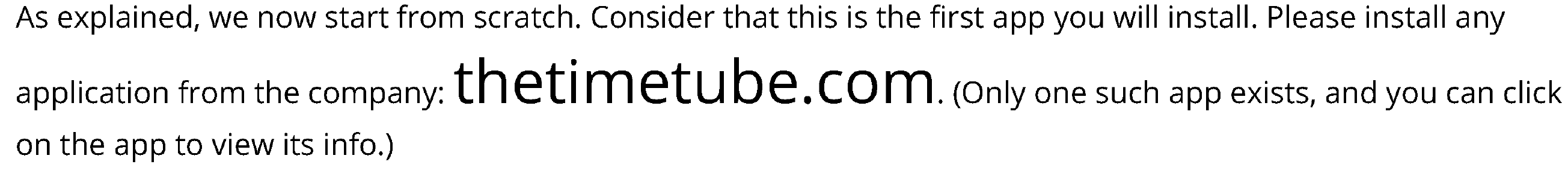}}
\subcaption{Module 1; Task $a$ : Vendor Familiarity}
\label{fig:exp-1-install}
\vspace{0.7\baselineskip}
\endminipage \quad\quad
\minipage{1\linewidth} 
\centering
\framebox{\includegraphics[width=0.96\linewidth]{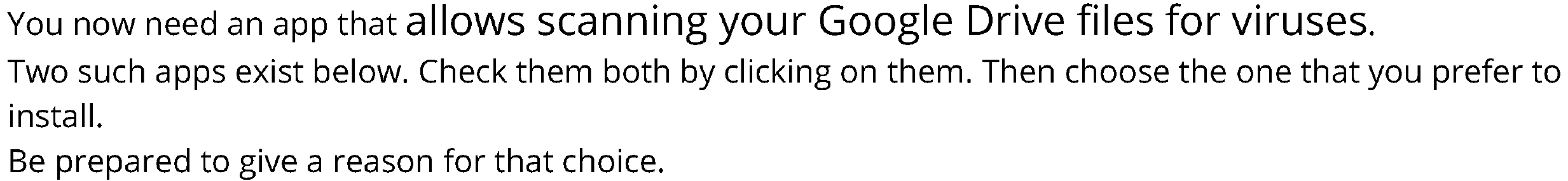}}
\subcaption{Module 1; Task $b$: App Selection}
\label{fig:exp-1-choice}
\vspace{0.7\baselineskip}
\endminipage\\
\minipage{1\linewidth} 
\centering
\framebox{\includegraphics[width=0.96\linewidth]{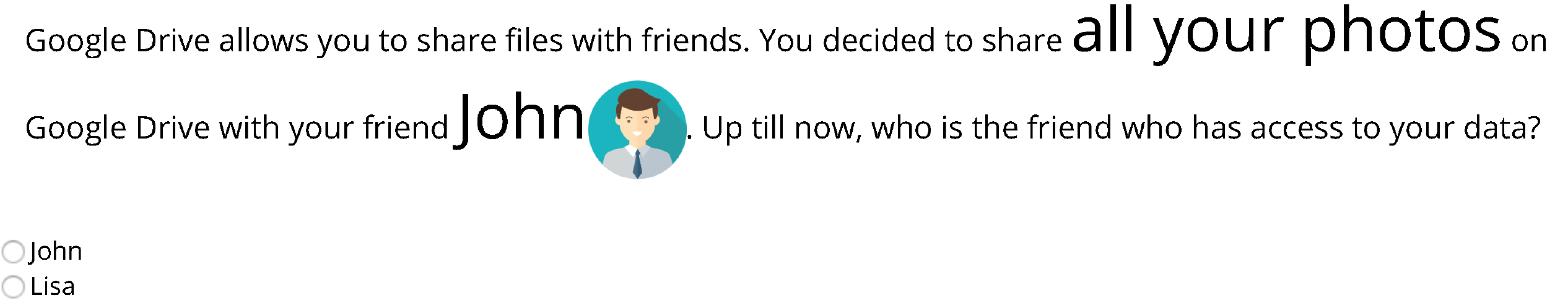}}
\subcaption{Module 2; Task $a$: Collaborator Awareness}
\label{fig:exp-2-intro}
\vspace{1.1\baselineskip}
\endminipage \quad \quad
\minipage{1\linewidth} 
\centering
\framebox{\includegraphics[width=0.96\linewidth]{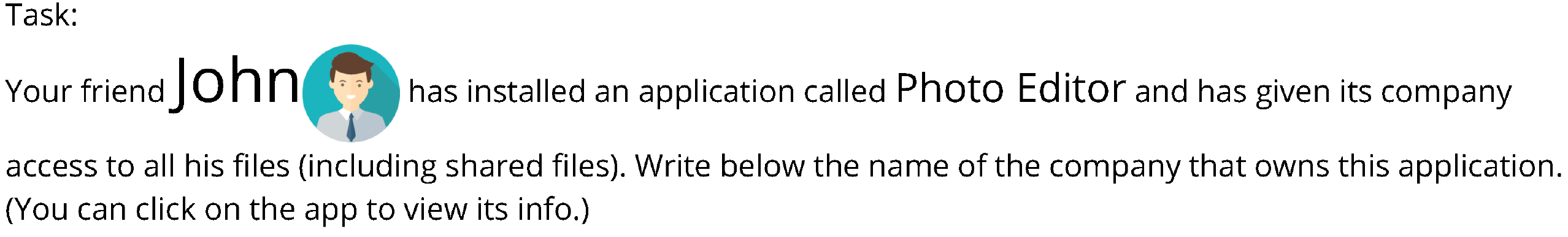}}
\subcaption{Module 2; Task $b$: Vendor Familiarity}
\label{fig:exp-2-install}
\vspace{0.7\baselineskip}
\endminipage\\
\minipage{1\linewidth} 
\centering
\framebox{\includegraphics[width=0.96\linewidth]{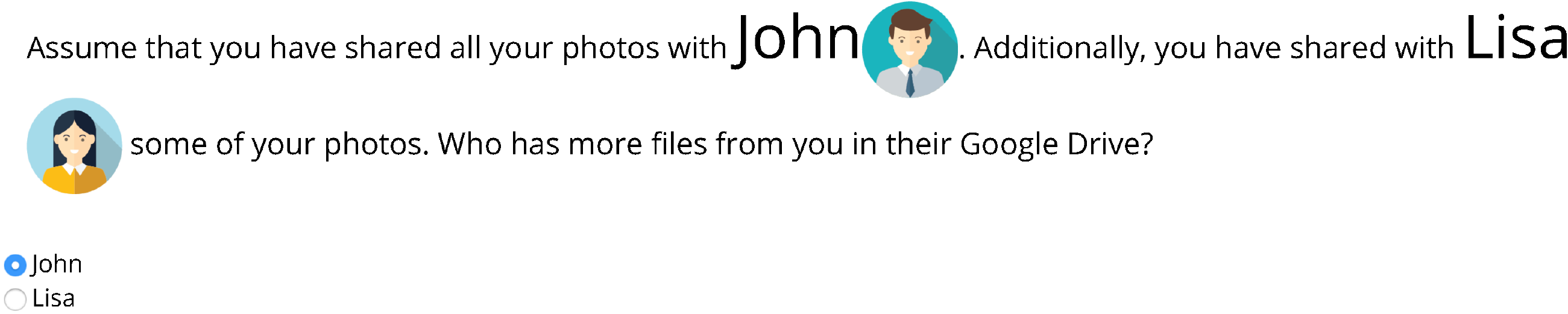}}
\subcaption{Module 3; Task $a$: Collaborator Awareness}
\label{fig:exp-3-intro}
\endminipage\\
\caption{Screenshots from the user study
}
\label{fig:screenshots}}
  \end{figure}

We now move to the detailed description of the modules and the results obtained.
These modules are summarized in Figure~\ref{fig:modules}, to which we refer henceforth.
We also show sample screenshots from the online study in Figure~\ref{fig:screenshots}.
The results are also presented in~Table~\ref{tab:studyResults}.

\tparagraph{Module 1 (Self-History Scenario)} tests whether the user is more likely to select an app from the same vendor she has just installed from before.
In step (a), the user is made aware the she installed an app from a specific \mbox{vendor $v$} (Figure~\ref{fig:exp-1-install}). 
In step (b), she is asked to install\footnote{Users were informed that this is a role-playing study, and no apps were actually installed.} an app that satisfies the given purpose (Figure~\ref{fig:exp-1-choice}) among a list of apps. Two of the listed apps were relevant, and one of them was from vendor $v$ itself. 

		Despite the participants being informed one step earlier that they installed an app from \enquote{thetimetube.com}, that did not make a difference in the \BL\ case: half of the users still chose the app from the new vendor \enquote{nitrosafe.org} (cf. Table~\ref{tab:studyResults}.  In the absence of traditional signals that users follow for deciding on apps (reviews, ratings, permissions), participants apparently made decisions that cancelled out, making the two apps equally favored across participants. \textbf{The vast majority of users were not approaching the installation from the angle of keeping their data with fewer shareholders}. Based on their provided justifications, they rather looked for other cues, such as selecting the app that, in their opinion, has a more comprehensive description, a more professional logo, a better sounding name, or a more trustable URL.
		Still, 12 users have explicitly mentioned in their text input that they chose an app \textit{because} it is from the same vendor they have dealt with earlier. Even then, neither of them has alluded to a privacy motivation behind the choice. These 12 participants mainly provided cross-app compatibility, interface familiarity, and satisfaction with the previous vendor as justifications. For example, one participant wrote: \enquote{\textit{I favoured Malware Scanner due to the fact that the name \enquote{thetimetube.com} was in the last app installed, and I tend to install apps from the same company due to cross-app compatibility usually found in apps by the same company.}}
		Interestingly, two users justified their installation of the app from the new vendor (nitrosafe.org) by writing that they had just installed an app from the same company before. This indicates that, \textbf{even when users try to account for previous decisions, they might find it difficult to remember the previous app vendors}. Given that our study had a short time span separating the current from the previous installation, we expect that such mistakes would be even more common in real scenarios when app installation instances are separated by longer time spans.
		
		The \HB\ group witnessed a much larger proportion of users who favored the option with less privacy loss. 72.2\% of the participants selected the app from the \enquote{thetimetube.com} (the vendor which already has access). The difference of 22.8\% compared to the \BL\ group is statistically significant (Fisher's exact test, $p\mbox{-value}=0.005$). Many of the participants who chose the app from \enquote{thetimetube.com} reported that they were motivated by the 100\% access that the app already has. We counted around 40 such users (\ie 57\% of the \HB\ group). Some of them went further and explicitly mentioned that their selection was motivated by giving data to fewer data owners (\ie more privacy). For example, one user wrote: \enquote{\textit{This company has access to all my files, so I would choose them as I don't want to have 2 companies with full access to my files}}. 
		
		In a nutshell, we were able to verify our hypothesis in this scenario: \textbf{the new privacy indicator leads users to more frequently choose the app from a vendor they already authorized}. Furthermore, we have discovered that the \textit{HB} Insights interface has indirectly made users think about various positive effects brought by using apps from the same vendor. This eventually lead them to make more privacy-preserving decisions.

\begin{table*}[ht]
\centering
\ra{1.3}
\begin{tabular}{@{}rcccccccccc@{}}
\toprule
\multirow{2}{*}{\textbf{Scenario}} & \phantom{a}& \multicolumn{2}{c}{\textbf{BL group}} & \phantom{a} & \multicolumn{2}{c}{\textbf{HB group}} & \phantom{a}   & \textbf{$\Delta$} & \phantom{a} & \textbf{$p$-value} \\ 
\cmidrule(lr){3-4} 
\cmidrule(lr){6-7}
 &  & \textit{VwA} & \textit{NV}  & & \textit{VwA} & \textit{NV}  &  & & & \\
\midrule
\\
 \textbf{Self History} &   & 50.0\%   &  50.0\%   & & 75.4\%   &  24.6\% & &  25.4\%  & &   0.003   \\
 \textbf{Collaborator's app} &  & 52.8\%   &  47.2\%  &  & 88.4\%   &  11.6\%  & &  35.6\%  & &   < 0.001   \\
 \textbf{Collaborator's vendor} &  & 58.3\%   &  41.7\%  &  & 82.6\%   &  17.4\%  & & 24.3\%  & &   0.002   \\
 \textbf{Multiple collaborators} &  & 44.4\%   &  55.6\%  &  & 82.6\%   &  17.4\%  & &  38.2\%  & &   <0.001   \\
\bottomrule
\end{tabular}
\caption{App selection statistics in the study; \textit{VwA}: vendor with access; \textit{NV}: new vendor. The comparisons in each experimental group were planned contrasts, and the $p\mbox{-values}$ of difference between the percentages of users who selected each app type were computed using Fisher's exact test}
\label{tab:studyResults}
\end{table*}

		\tparagraph{Module 2.1 (Collaborator's App Scenario)} tests the likelihood that the participant selects the same app that her collaborator had used.
		 In step (a), the participant is made aware that she had shared all her photos with a friend $f$ (Figure~\ref{fig:exp-2-intro}). 
		 For more familiarity, we also added a picture for each of the two fictitious friends throughout the study. In step (b), the user is made aware that her friend $f$ has installed an app $a_0$ (Figure~\ref{fig:exp-2-install}) 
		 from vendor $v$. She is asked to type the name of the app's vendor (\enquote{paste} option was disabled in the input field to further ensure the participant is aware of the vendor). In step (c), the user is asked to install an app with a certain purpose (similar to Figure~\ref{fig:exp-1-choice}).
		 One of the two matching apps is app $a_0$.
		
		Similar to the previous module, the \BL\ group witnessed an almost even split between \enquote{Online Player}, installed previously by the friend, and \enquote{Enjoy Music Player}, from a new vendor (cf. Table~\ref{tab:studyResults}. We also noticed that 20 participants in this group justified their decision by mentioning that their friend has used the app. Still, neither of them alluded at privacy reasons in their justifications. Instead, the two most prevalent motivations were (1) considering the friend's use of the app as a \textit{recommendation} or (2) achieving \textit{compatibility} with their friends' app, which facilitates data sharing within the app itself. Quoting one user: \enquote{\textit{This is the same app my friend is using so it should be quite compatible for us to both share.}} 
		
		In addition to having a significant 35.6\% difference in the case of the \HB\ group,
		we noticed that 32 users mentioned the existing data access as a reason for choosing the app \enquote{Online Player}. Also, 26 users referred to the fact that the friend has installed this app before (including those who mentioned both of the previous reasons). Unlike the \BL\ group's justifications though, where the friend's recommendation and the app's compatibility prevailed, the privacy issue was explicitly brought up by at least 10 users. One participant put it as follows: \enquote{\textit{Thanks to John, they have already access to 70\% of my data. Sharing the last 30\% isn't as bad as sharing 100\% of my data with driveplayer.com.}}

		\tparagraph{Module 2.2 (Collaborator's Vendor Scenario)}
		 We proceed in steps (d) and (e) as in the previous scenario's steps (b) and (c), with the difference that a \textit{new} app from $v$ is included among the options in step (e) instead of the exact same app $a_0$.
		One interesting insight from this scenario is that \textbf{the line between the company and the app is blurred in the minds of several users} who used the two entities interchangeably. In fact, 3 users in the \BL\ group and 7 participants in the \HB\ group justified their choices by mentioning that their friend installed the \textit{same app} before, which was not the case. For example, one user wrote: \enquote{\textit{this app already has access to my files, and I don't want to install any new app.}}

		\tparagraph{Module 3 (Multiple Collaborators Scenario)} Given collaborators $f_{more}$ and $f_{less}$, where the user shares much more data with $f_{more}$, this scenario checks the likelihood of the participant authorizing an app that $f_{more}$ has installed.
		In step (a), the participant is made aware that $f_{more}$ has access to more data than $f_{less}$ (Figure~\ref{fig:exp-3-intro}). 
		In steps (b) and (c), the participant is made familiar with the apps each of the friends installed (similar to Figure~\ref{fig:exp-2-install}).
		In step (d), the user is asked to select an app with a specific purpose. The two friends' apps are the only ones matching, and the choice is to be made between them (similar to Figure~\ref{fig:exp-1-choice}).

		In the \BL\ group, we had 44.4\% of the participants choosing the app installed by $f_{more}$. Still this percentage is relatively close to an equal split between the two apps. Out of this percentage, 13 users justified their choice by mentioning that they were encouraged to follow the choice of friend $f_{more}$. Even though they did not mention privacy, the larger number of files shared with $f_{more}$ was often used as a justification. For example, one participant wrote:
		\enquote{\textit{This is the app that John already uses, and he has access to all of my files. The PDF Mergy app is used by Lisa, but she only has access to part of my files.}}

		In the \HB\ group, around 82.6\% chose the app previously installed by the friend $f_{more}$, which is significantly more than those in the \BL\ case (Fisher's exact test, $p\mbox{-value}<0.001$). Looking at the justifications, around 37 users explicitly mentioned the higher access level that this app already possesses as a reason for their choice. Privacy was additionally mentioned by 8 of these users. Quoting one of them: \enquote{\textit{PDF Mergy already has access to 70\% of my files. Using PDF Files Merger would unnecessarily increase third party app access to my files.}} However, we still had 2 users who went for the app with less existing access, with one of them saying he favors the app that only \enquote{\textit{had accessed 30\% of files before installation}}. What was interesting though is that \textbf{almost all users who mentioned friends were actually making a comparison between the two friends' existing access level}, regardless of their final choice.
	
		\subsection{Concluding Survey}
			At the end of the user study, users were presented with a final set of questions. 
			We asked them whether they would like to be notified when a friend installs an app that gets access to their shared files. 
			Around 92\% of users in the \BL\ group and 90\% of users in the \HB\ group agreed. We further asked the participants whether they are fine with a collaborator being notified when they install applications that access files shared with that collaborator. The percentage of people who agreed dropped to 75\% in the \BL\ and 78\% in the \HB\ group. The relatively small difference between the answers to these two questions highlights that \textbf{only a minority of users is not willing to make the trade-off of contributing to the overall system.} Such users can be given the option to not use privacy indicators based on their friends' decisions.
            
			Next, users were asked the following question \textit{``Assume you have installed an application called YouMusic from a company called Musicana and gave it access to all your files on Google Drive. Now you are considering installing an application called YouVideo from the same company. How do you think that this application will affect your privacy:''}. Only 11\% of each group replied by \textit{``negatively''}. The vast majority in both groups either perceived the avoidance of a new vendor as a positive outcome or considered that the privacy loss will remain the same. Interestingly, the users in the \BL\ showed a similar reasoning in justifying their choices as the \HB\ group although the latter were primed about these aspects via the privacy indicators. This indicates that \textbf{the privacy indicators actually match the first intuition for a large fraction of users}.

		\subsection{Discussion and Limitations}
		
		Overall, we found out that, in the three modules, participants in the \HB\ group were significantly more likely to install the app with less privacy loss (\ie the app from the vendor with the largest share of the user's files) than those in the \BL\ group. 
		Despite showing the efficacy of History-based Insights, our study still has its limitations. In order to get a large, diverse sample size, we resorted to a web experiment based on role-playing with hypothetical data. It would be interesting to see how such results extrapolate to the case where users' own data is in question. 
        
		Moreover, in our design, we have abstracted several factors (e.g., ratings and reviews), which have been previously studied in similar ecosystems~\cite{Kelley:2013}, in order to focus on one factor. These factors might have diluted the effect of the privacy indicator. Still, we conjecture that, although the absolute values of our findings might not strictly apply, the differences between the two groups will still be practically significant. 
        
		Additionally, in this paper, we have investigated only one type of history-based privacy indicators. Evidently, such indicators can be integrated at different stages of the app installation process. For example, they can be part of the recommendation strategy for suggesting alternative apps. They can also be included in the apps' search interface. Apps can also be labelled as ``privacy preserving'' in the web store based on this metric. It is also possible that the privacy indicator is only shown when the vendor has existing access to the user's data. This might serve to reduce the habituation effect and the information overload. The best choice among these deployment scenarios needs further investigation.
        
		Furthermore, it is important to note that, although our experimental interface mentions the collaborators' name in the explanation under the progress bar, this does not have to be the case in actual deployments. We hypothesize that removing the name will not have a significant impact on the results as it was not highlighted in the interface. This allows the CSP to relay such information to the users without exposing sensitive data about particular collaborators. The CSP can resort to more sophisticated anonymization methodologies, such as showing a non-exact percentage that can be mapped to multiple collaborators. Exploring the impact of these techniques is left for a future work. Moreover, we note that this anonymization might not be needed at all in the enterprise settings, where apps installed by team members are supposed to be visible for the administrators. As we show in Section~\ref{sec:teams}, a significant reduction in privacy loss can be achieved without even accounting for decisions by users external to the team.
		
        Finally, the privacy indicator in our study has addressed two granularity levels: full and per-file access. However, the same indicator can be extrapolated to the case of per-type access. For example, the interface can say: ``The app's company already has access to 70\% of your \textit{photos}'' (instead of \textit{files}).

		\section{Large Networks' Simulations}
		\label{sec:simulation}

		In the previous section, we showed the significant change that our privacy indicator can effect through encouraging users to make History-based decisions. We will tackle the next research question, where we investigate the impact of adopting such privacy indicators on the privacy risk in realistic scenarios with large user networks.
		As we are not in the position of the CSP to study an actual implementation of the \HB\ Insights interface over time, we will perform a simulation of potential users' installation behavior. 
		We will base this on both the crowdsourced decision model inferred from the user study and on new collaboration networks that we construct.

		\subsection{Simulation Data}
		\label{sec:simData}
		
		\tparagraph{Collaboration Networks}
		For the purposes of this simulation, we constructed the following three networks:
		\begin{itemize}[leftmargin=*]
			\item
			\textbf{Inflated Google Drive Network}: We used the standard degree-driven approach for network topology generation to construct a larger Google Drive network based on the one in the~\textit{PrivySeal Dataset} of Section~\ref{sec:privsealDataset}~\cite{mihail2002generating}. Based on an input user degrees' distribution from that dataset, we particularly used the \textit{Configuration Model} as described by Newman~\cite{newman2003structure} and implemented by the library NetworkX~\cite{schult2008exploring} for inflating the graph. This model generates a random pseudograph (a graph with parallel edges and self-loops) by randomly assigning edges to match an input degree sequence. 
			We removed the self-loops and parallel edges a posteriori from the generated graph. 
			In the end, we had a collaboration graph with 18,000 users and 138,440 edges. This graph is, by construction, a connected graph, with an average node degree of 15.
			\item
			\textbf{Paper Collaboration Network}: In an effort to have a realistic, large collaboration network without resorting to graph inflation, we relied on the Microsoft Academic Graph, which consists of records of scientific papers along with the authors and their affiliations~\cite{Microsoft}. We used a snapshot of 50,000 papers, and we constructed the collaboration graph based on it. We ended up with 41,000 collaborators and 199,980 edges. The graph itself is not a connected graph but is rather constructed of around 1700 connected components. The average node degree is 4. Our rationale is that this graph captures a realistic scenario of users collaborating on authoring documents, which is, in fact, an activity achieved via cloud services nowadays. Hence, it is fit for showing the efficacy of our privacy indicators.
			\item
			\textbf{Team Collaboration Network}: We used the same academic graph in order to construct a network of teams. A team is defined as a frequently collaborating group of people. Motivated by research around community detection~\cite{Malliaros201395}, we use Strongly Connected Components (SCCs) in order to label teams in our graph. We ended up with 16,400 users split over 1700 teams. Unlike the previous two networks where users themselves are the data subjects (whose privacy is to be optimized), members of each team in this network consider their team as the data subject.
			
		\end{itemize}

		\tparagraph{Sharing and Installation Patterns}
		In order to closely model the user characteristics in Google Drive, we assigned to each user in the collaboration networks a file sharing distribution and a number of apps corresponding to a user with a matching degree in the PrivySeal Dataset.

		\tparagraph{Apps}
		As we wanted to perform the simulation with a much larger number of users than we had in the dataset described in Section~\ref{sec:privsealDataset}, we also needed a larger collection of apps. 
		Given that Google Chrome Store has only around 500 apps that are tagged by the \enquote{Works with Google Drive} tag, we decided to also include all Google Chrome Apps in the dataset (\ie even those that do not have this tag). As far as the simulation is concerned, this step is justified since the only  realistic information that we will rely on is the distribution of vendors per app. It is fair then to assume that this distribution does not differ significantly between the general category and the Google Drive category.
		Hence, we augmented the PrivySeal Dataset via apps from the Google Chrome Store to arrive at 1000 apps. In addition to the app's installation count and vendor name, we also collected the set of \enquote{\textit{Related Apps}} that the store displays for each app. This is because, in our simulation, we will assume that users have the choice to choose the app itself or one of its related apps. Again, this is a fair assumption as these related apps are mostly the apps which deliver a close functionality to the app itself, and we will only rely on them to model the alternatives at each simulation step.

		\tparagraph{User Decision Models}
		For the purpose of this simulation, we define 3 user decision models:
		\begin{itemize}[leftmargin=*]
			\item 
			\textbf{Fully Aware Model (\FA): } the user always makes the decision that minimizes the privacy loss of the data subject, taking into account all previous installation decisions by her and by her collaborators.
			\item 
			\textbf{Experimental History-based Model (\EHB)}: the user takes decisions similar to what a random user of the \HB\ experimental group does. 
			In specific, we model those users as taking a history-based decision with probability $q$ and making a random app choice with probability $1-q$. We set $q$ based on the number of users who mentioned the app' existing access in \textit{writing} as a reason for their choice in each module of Section~\ref{sec:study}. Based on Module 1's users' responses, we set $q=0.57$ when the user encounters a vendor she previously authorized. Based on Module 2, we set $q=0.70$ whenever the user is presented with one vendor previously authorized by a single collaborator. Based on Module 3, we set $q=0.67$ for the cases where the user is presented with multiple vendors previously authorized by her collaborators. In all of these cases, the user will select the vendor with the minimal resulting \textit{Aggregate} $\VFC_u$ with probability $q$.
			\item
			\textbf{Experimental Baseline Model (\EBL): } the user takes decisions similar to what a random user of the \BL\ experimental group does.
			As users in practice are rarely informed of what their friends have installed before, we do not integrate this knowledge into the model. Hence, we only account for the case of Module 1, where the user's previous decisions are concerned. Based on the fraction of users who mentioned the app's existing access as a motivation for their choice, we set the probability of taking history-based decision in this model as $q=0.18$.
			
		\end{itemize}

		In the special case of the team collaboration network, users who take history-based decisions account for their own decisions and the decisions of their team members only. We do not consider that users account for decisions taken by members of other teams. This is to demonstrate the potential of the privacy indicators under strict conditions.

		\subsection{Simulation Details}

		\begin{algorithm*}[t]
			\caption{Simulation Steps}
			\label{alg:simulation}
			\begin{algorithmic}[1]
				\State Initialize $\VFC_u$ value to 0 for each user
				\For{t $\gets$ 0 to N} \Comment{$N$ is total number of steps}
				\State select a random user $u_0$ based on user's app installation frequency\label{lst:line:userSelect}
				\State select a random new app $a_0$ based on app's installation count\label{lst:line:appSelect}
				\State	$A_{rel}:=  \{a_0\} \cup \mbox{(set of related apps of } a_0)$ 
				\State	$V_{rel}:= $ set of vendors of apps in $A_{rel}$
				\State $r:= $  a random rational number in the range [0,1]
				
				\If{user had installed apps by vendors in $V' \subset V_{rel}$}
				\If {($r<q$\textit{(group,\enquote*{same vendor}}))} \Comment{$q$ is a function of the user decision model; $group$ is the experimental group}
				\State select a random vendor $\hat{v}\in V'$ 
				\State install the app $\hat{a}$ in $A_{rel}$ from vendor $\hat{v}$ 
				\Else
				\State	install app $a_0$
				\EndIf	
				\ElsIf {$\exists$ ($c \in C(u_0)$ who installed apps by vendors in $V' \subset V_{rel})$} 
				\State compute $\VFC_{u_0}(\{v\})$ for each vendor $v$ in $V'$ at this time step
				\State select the vendor $\hat{v}\in V'$ with highest $\VFC_{u_0}(\{v\})$ at this time step
				\If {($r<q($\textit{group,\enquote*{collaborator vendor}}))}
				\State install the app $\hat{a}$ in $A_{rel}$ from vendor $\hat{v}$ 
				\Else
				\State install app $a_0$
				\EndIf	
				\Else
				\State install app $a_0$
				\EndIf
				\For{all $u \in \{u_0\} \cup C(u_0) $}
				\State update \textit{Aggregate $\VFC_u$} for $u$ \Comment{recompute it via Equation~\ref{equation:AggVFC}}
				\EndFor
				\State update the average \textit{Aggregate} $\VFC$ over all users
				\EndFor
			\end{algorithmic}
		\end{algorithm*}
        		
		We now move to the description of the simulation itself, which is detailed in~Algorithm~\ref{alg:simulation}. 
		We had three simulation groups, named after the three decision models: \FA\ group, \EHB\ group, and the \EBL\ group. 
		The simulation was run until the average number of apps installed across by users reached 30 apps\footnote{Comparatively, mobile users have accessed 26.7 smartphone apps on average per month in the fourth quarter of 2014~\cite{Nielsen}.}. On a high level, at each simulation step, the following actions are performed:
		\begin{itemize}[leftmargin=*]
			\item
			A user is selected from the collaboration network via a weighted random sampling based on the assigned app installation frequencies (line~\ref{lst:line:userSelect}). 
			This accounts for the diversity of users' installation frequencies.
			An app $a_0$ is selected from the simulation apps' dataset via a weighted random sampling based on the actual app installations count in Google Chrome Store (line~\ref{lst:line:appSelect}). 
			That way, popular apps are installed more frequently (as is the case in practice).
			\item 
			A user decision is simulated. The user is assumed to be choosing the app $a_0$ or one of its related apps. This choice is made depending on the user's decision model, as explained previously. 
			\item
			Finally, the average \textit{Aggregate} $\VFC$ is computed based on all users' \textit{Aggregate} $\VFC_u$.
			
		\end{itemize}

		\subsection{Simulation Results}
		\label{sec:simResults}
		
		\begin{figure*}[t!] 
			\centering
            
			\minipage[b][][b]{0.22\textwidth} 
			\includegraphics[width=1\linewidth]{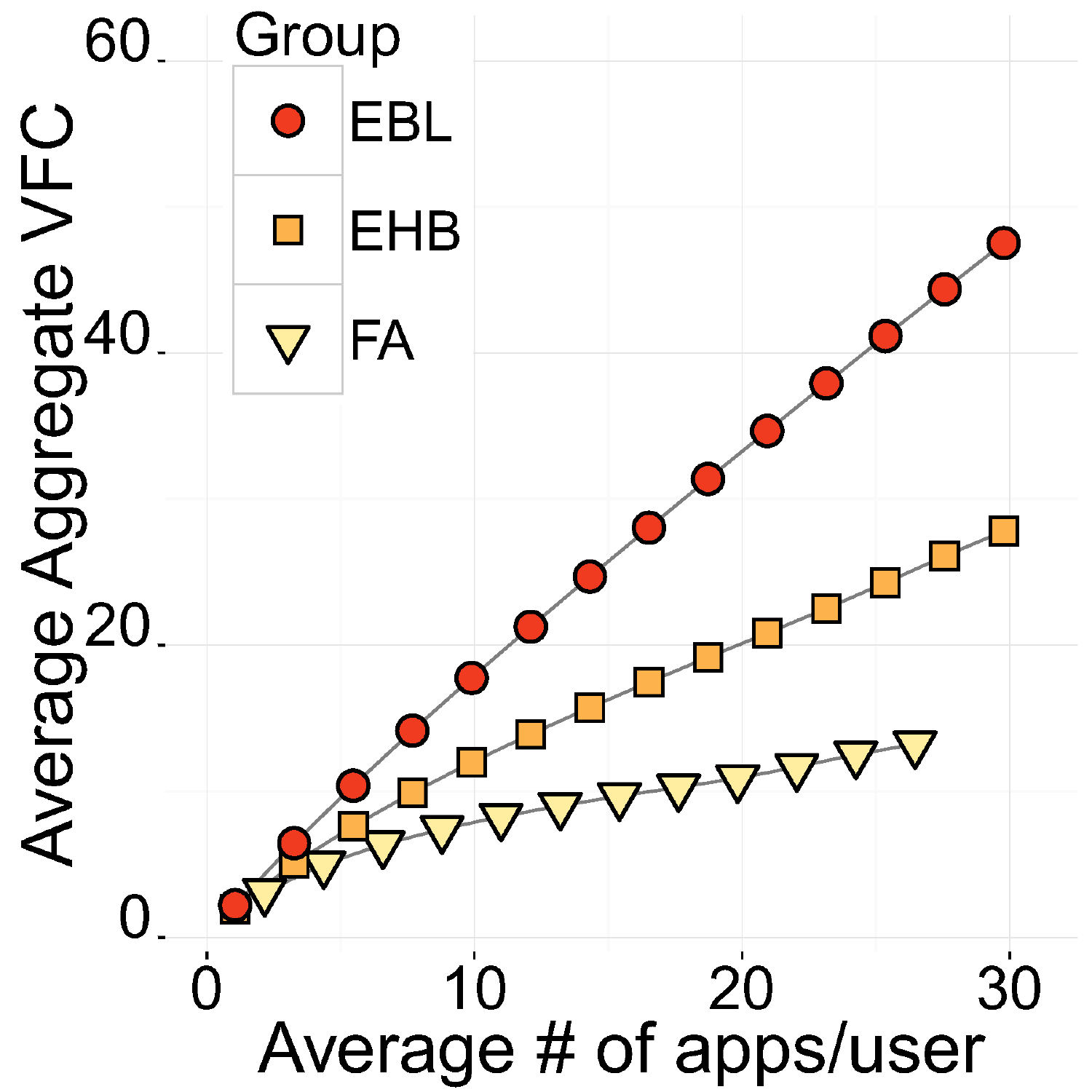}
            \subcaption{$\VFC$ evolution}
			\label{fig:simulation_groups_inf}
			\endminipage\quad
			\minipage[b][][b]{0.22\textwidth} 
			\includegraphics[width=1\linewidth]{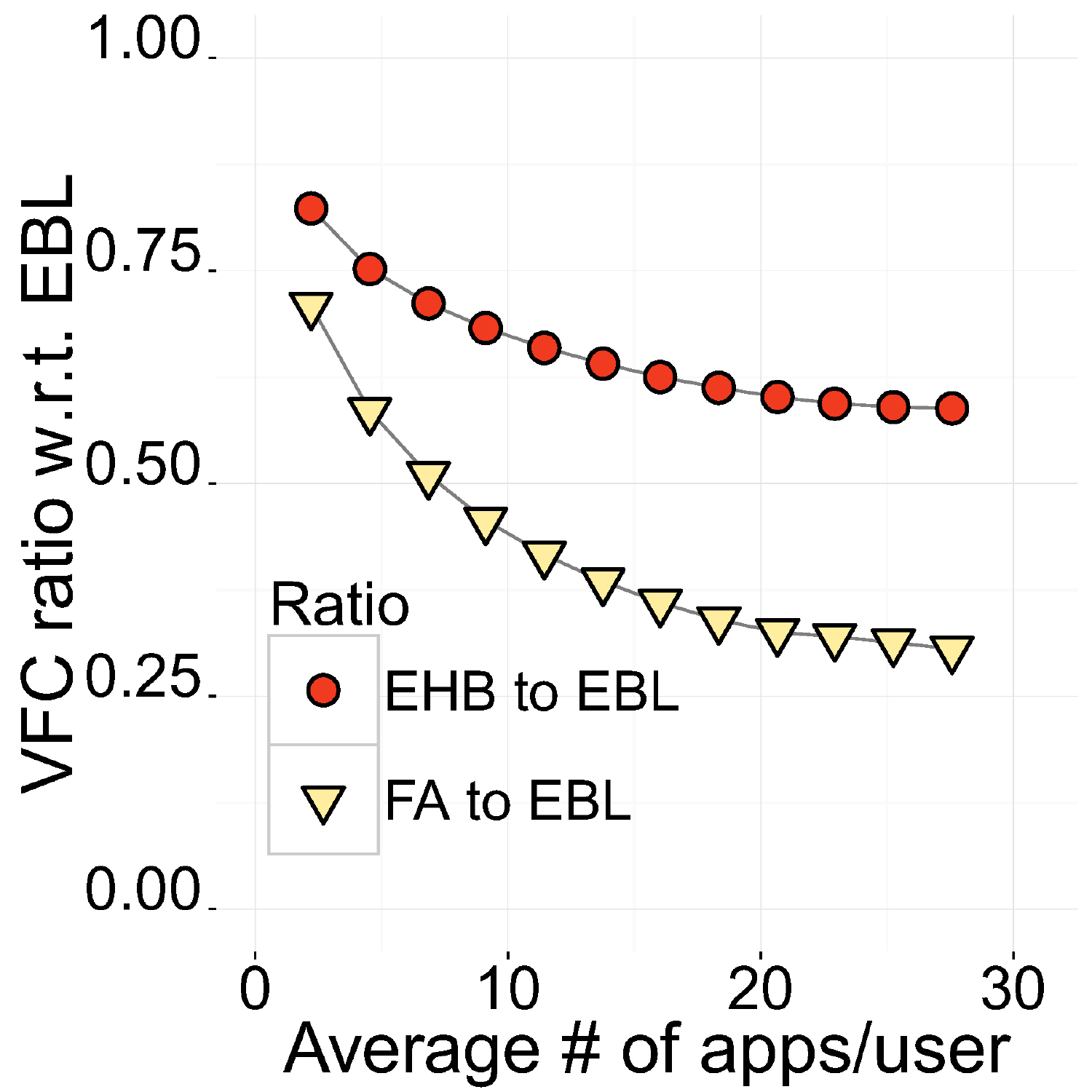}
            \subcaption{Evolution of the ratios w.r.t. the $EBL$ group}
			\label{fig:ratio_inf}
			\endminipage
			\minipage[b][][b]{0.51\textwidth} 
			\includegraphics[width=1\linewidth]{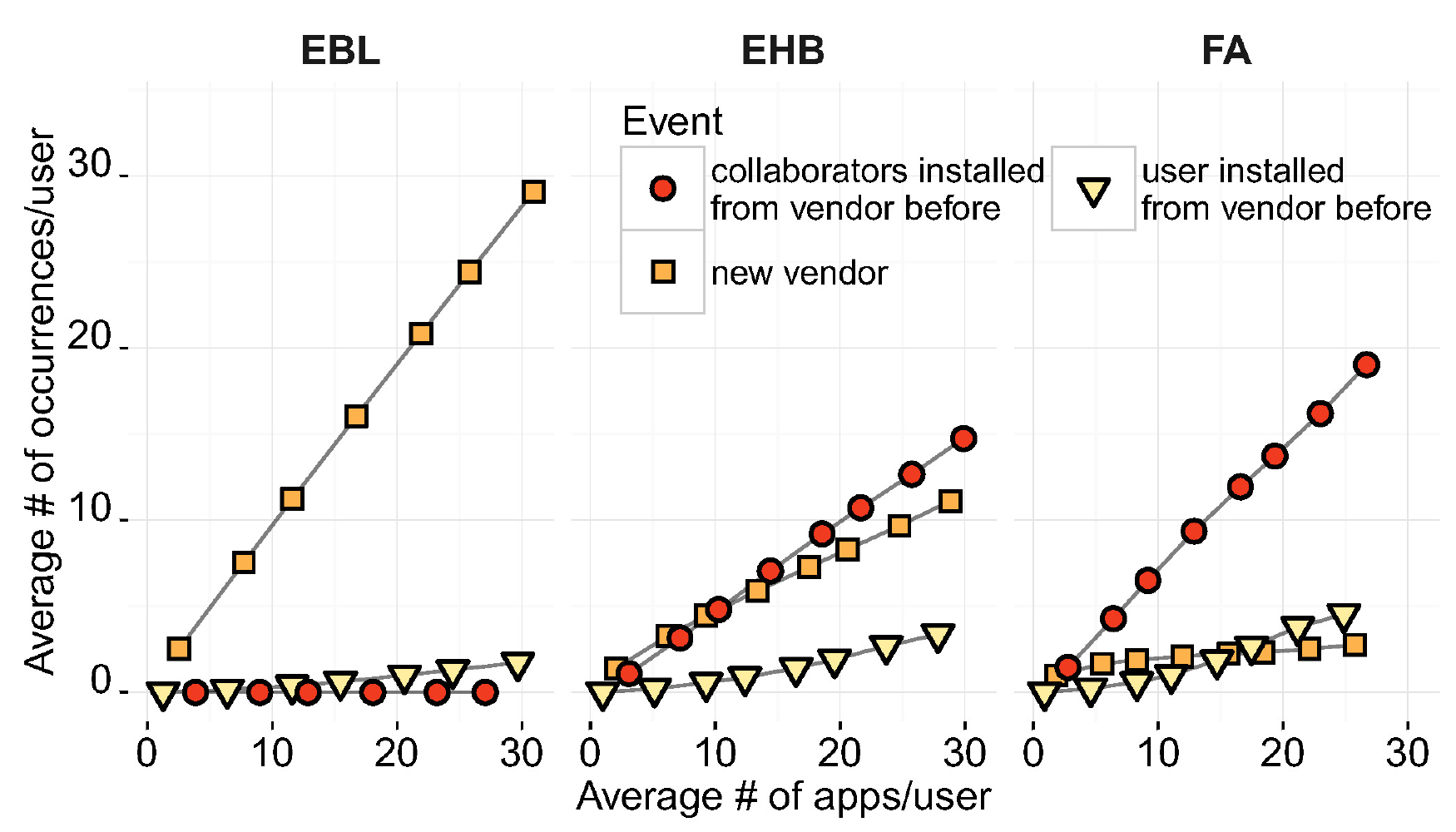}
			\subcaption{Evolution of the types of installation events}
			\label{fig:events_inf}
			\endminipage\\
			\caption{Simulation results in the \textbf{inflated network}}
            \vspace{1\baselineskip}
			\minipage[b][][b]{0.22\textwidth} 
			\includegraphics[width=1\linewidth]{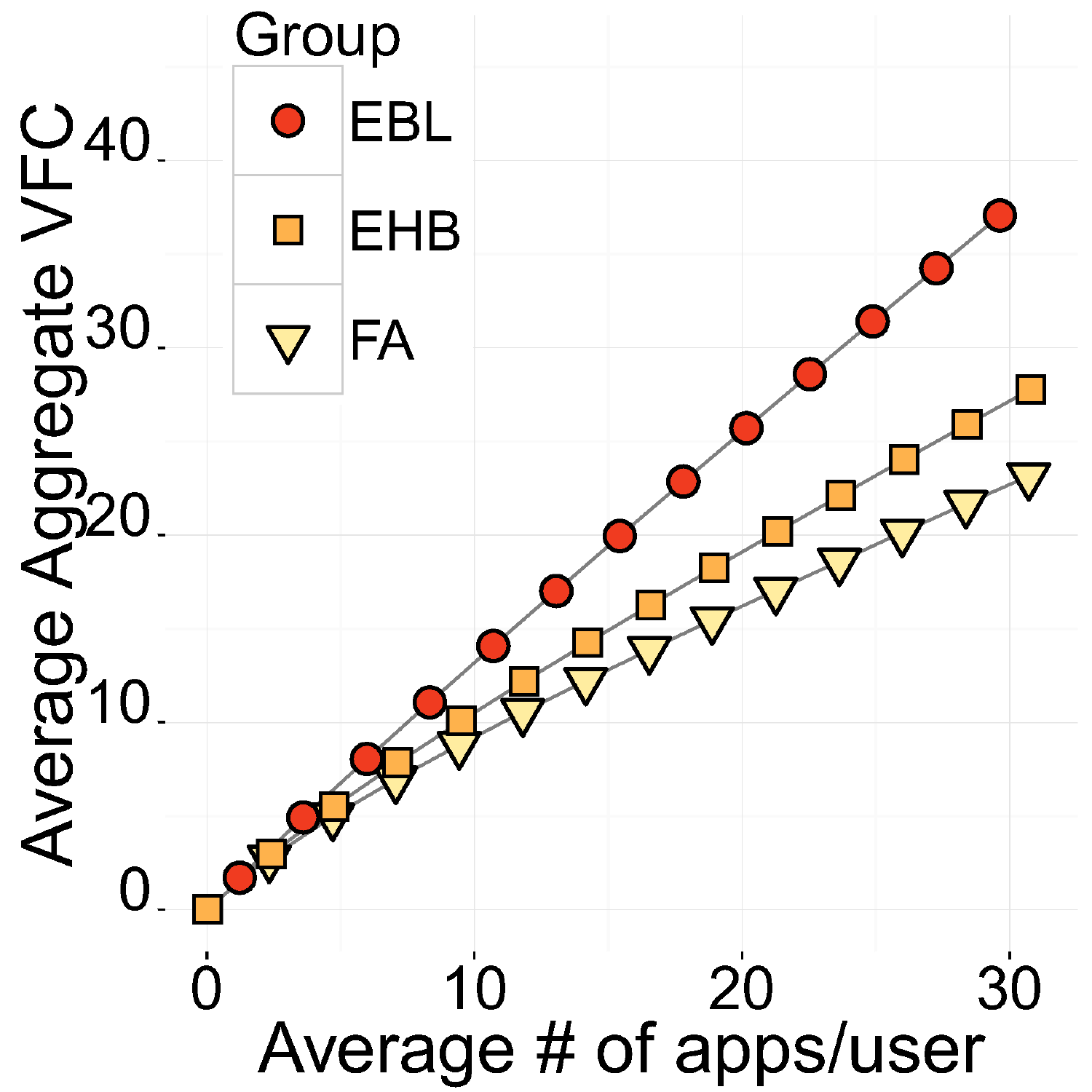}
			\subcaption{$\VFC$ evolution}
			\label{fig:simulation_groups_auth}
			\endminipage 	
            \minipage[b][][b]{0.22\textwidth} 
			\includegraphics[width=1\linewidth]{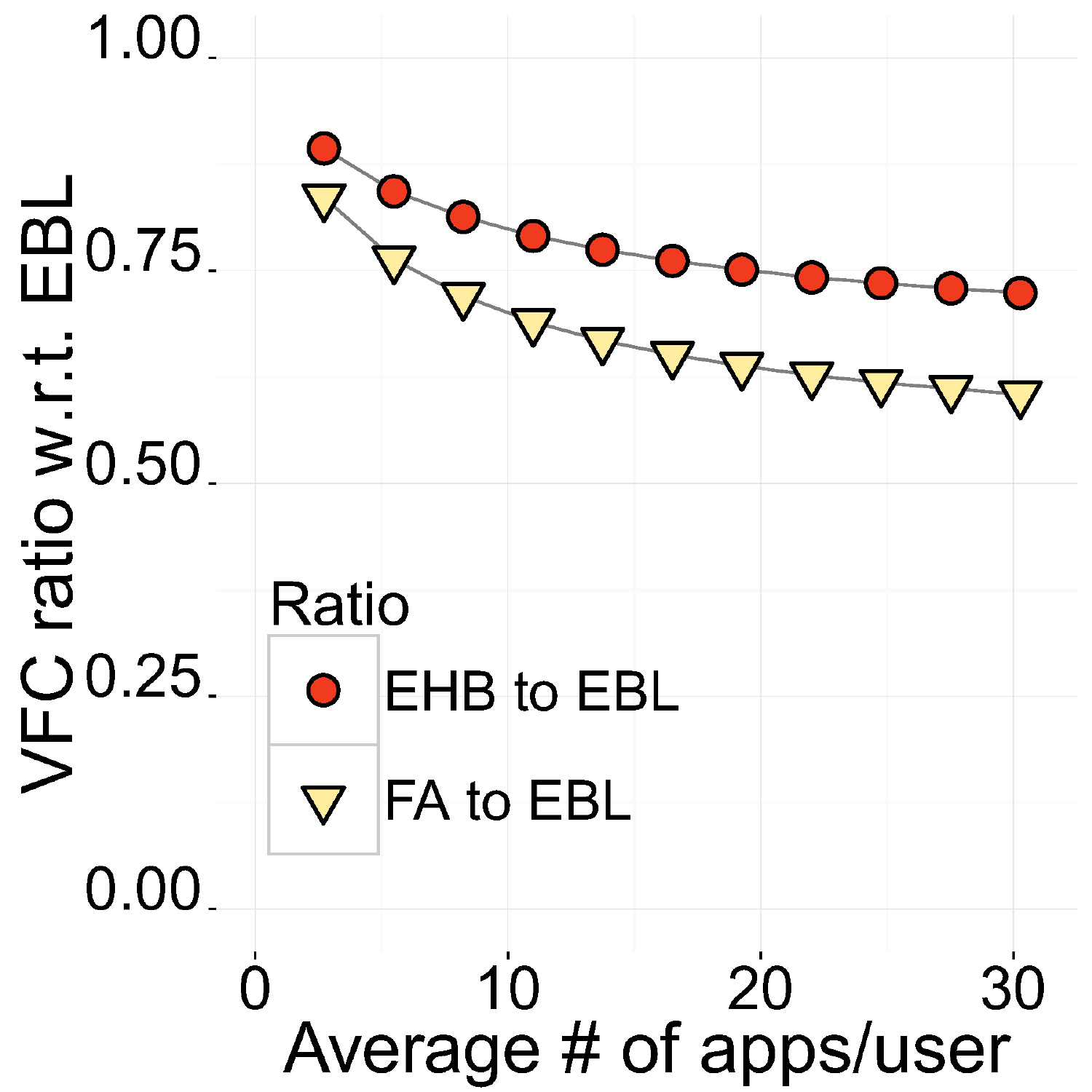}
            \subcaption{Evolution of the ratios w.r.t. the $EBL$ group}
			\label{fig:ratio_auth}
			\endminipage
            \minipage[b][][b]{0.51\textwidth} 
			\includegraphics[width=1\linewidth]{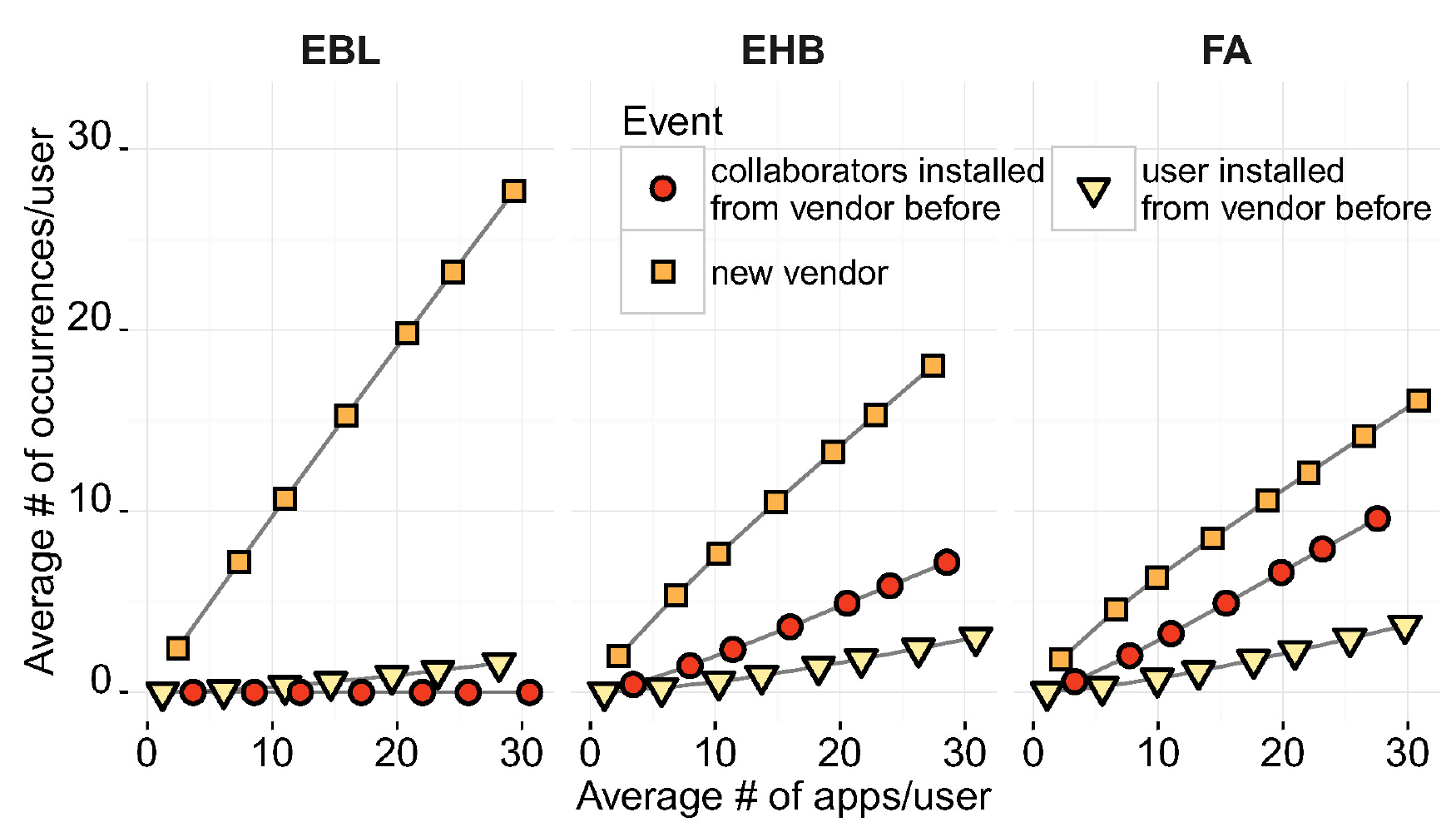}
			\subcaption{Evolution of the types of installation events}
			\label{fig:events_auth}
			\endminipage\\
			\caption{Simulation results in the \textbf{author-based network}}
            \vspace{1\baselineskip}
			\minipage[b][][b]{0.22\textwidth} 
			\includegraphics[width=1\linewidth]{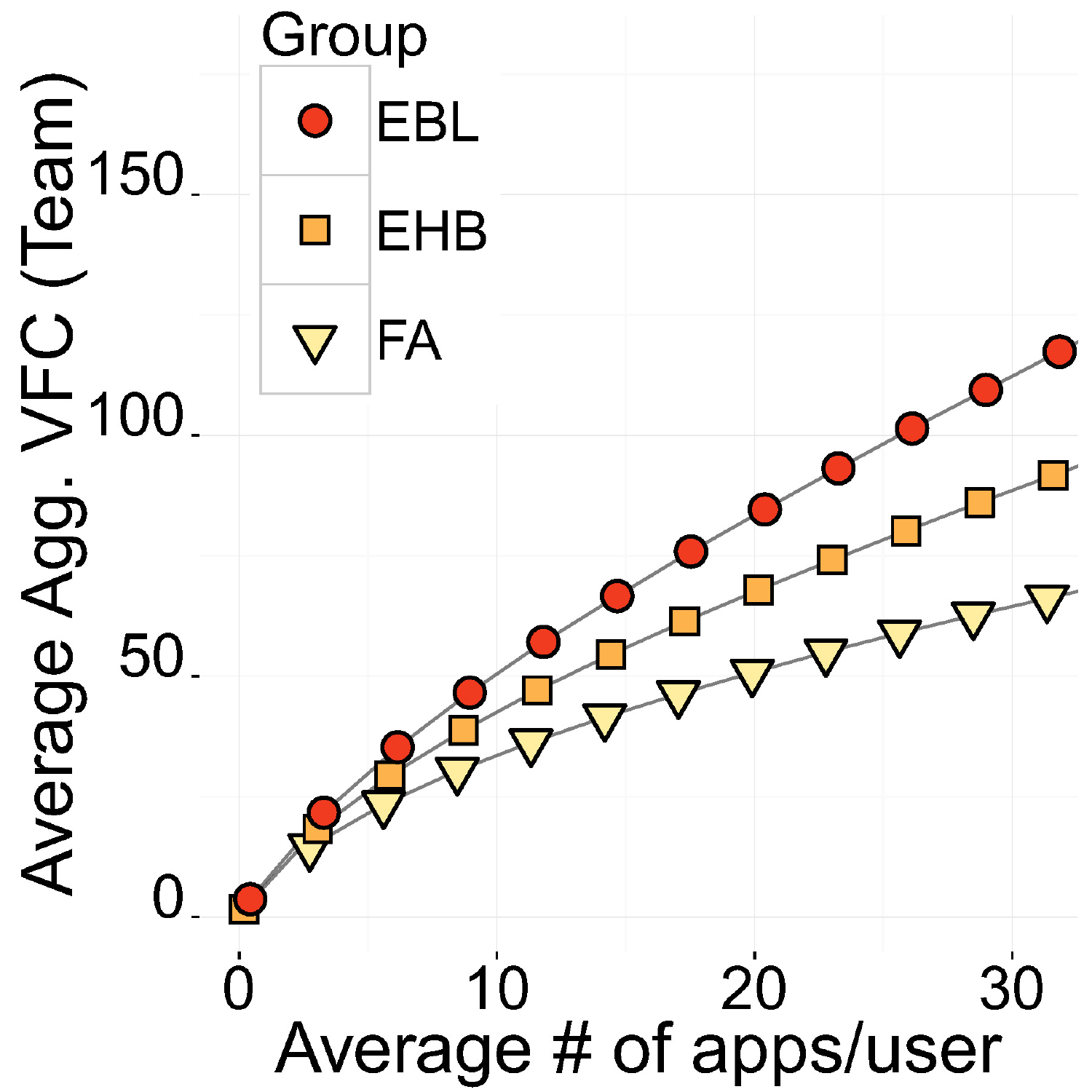}
			\subcaption{$\VFC$ evolution}
            \label{fig:simulation_groups_team}
			\endminipage
 			\minipage[b][][b]{0.22\textwidth} 
			\includegraphics[width=1\linewidth]{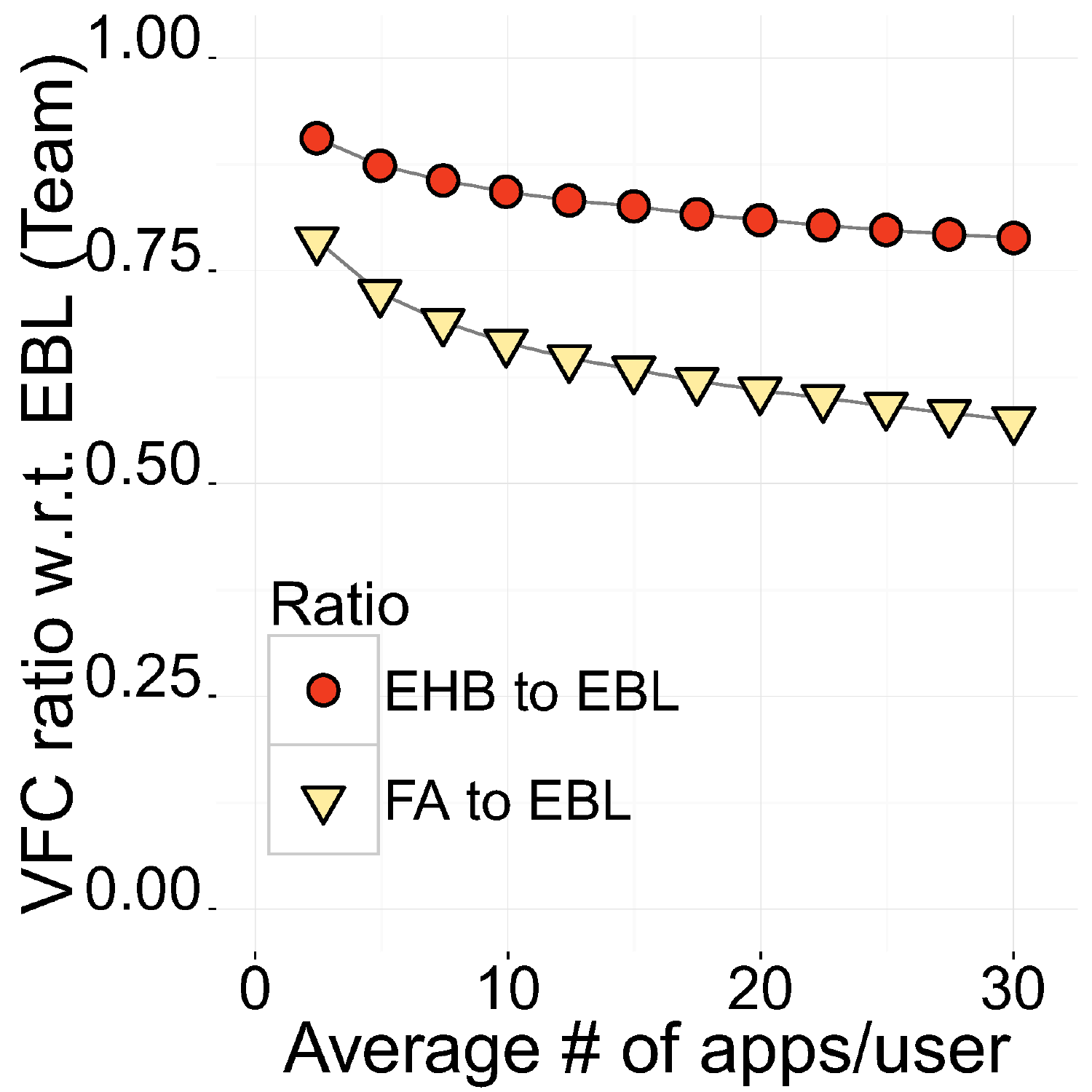}
            \subcaption{Evolution of the ratios w.r.t. the $EBL$ group}
      		\label{fig:ratio_team}
			\endminipage
			\minipage[b][][b]{0.51\textwidth} 
			\includegraphics[width=1\linewidth]{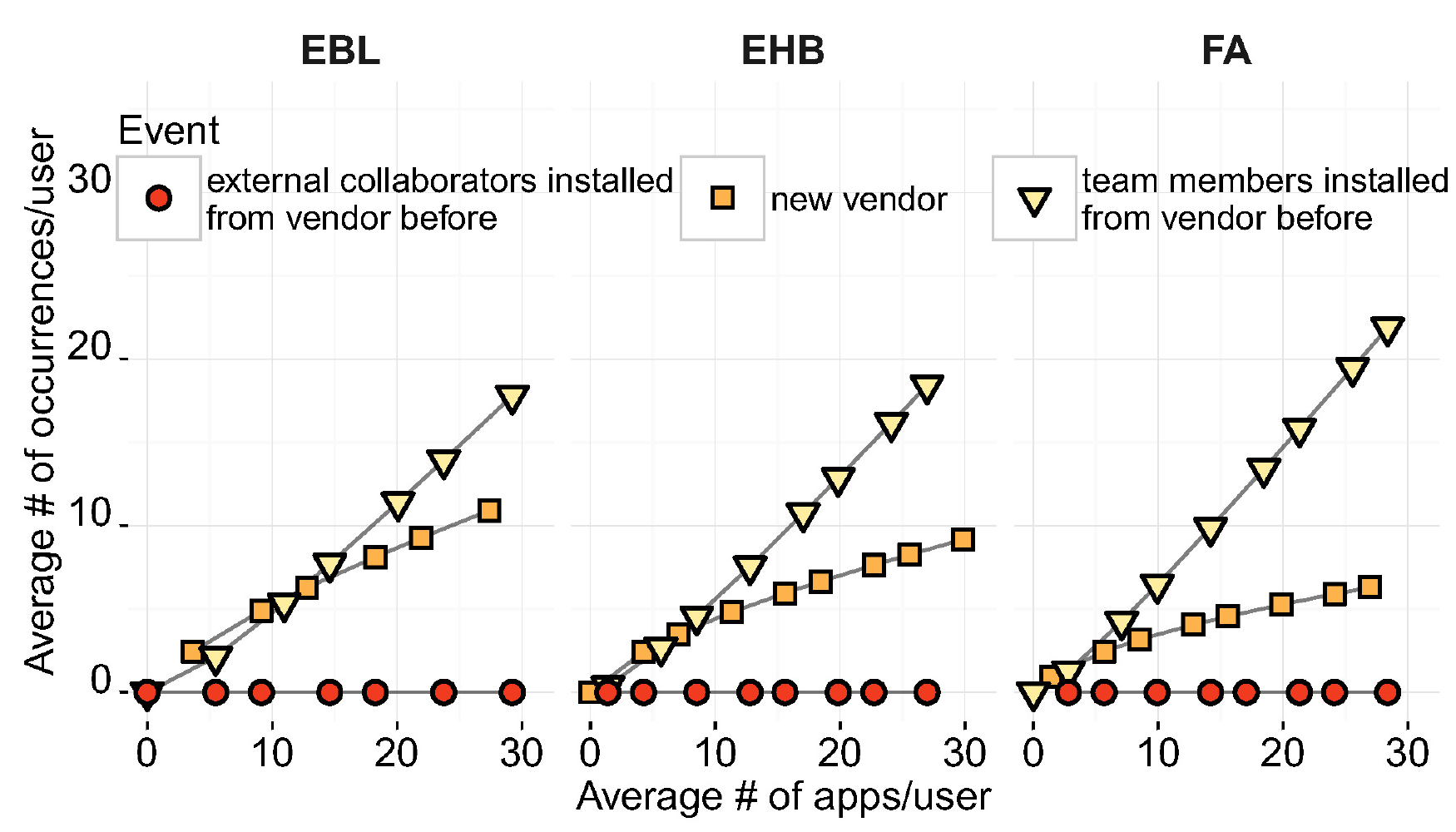}
			\vspace{-\baselineskip}
			\subcaption{Simulation results in the teams' network}
			\label{fig:events_team}
			\endminipage
			\caption{Simulation results in the \textbf{teams' network}}
            \label{fig:sim_team}
           
		\end{figure*}

		To demonstrate the simulation results, we show three types of figures per collaboration network. On a high level, in Figures~\ref{fig:simulation_groups_inf},~\ref{fig:simulation_groups_auth}, and~\ref{fig:simulation_groups_team},
		we show how the privacy loss (quantified using the average Aggregate-$\VFC$) in each group evolves as users install more apps. In Figures~\ref{fig:ratio_inf},~\ref{fig:ratio_auth}, and~\ref{fig:ratio_team}, 
		we show ratios of the privacy loss in the two experimental groups \EHB\ and \FA\ with respect to the baseline \EBL\ group. Finally, Figures~\ref{fig:events_inf} and~\ref{fig:events_auth}, and~\ref{fig:events_team} 
		show the actual events contributing to the privacy loss growth, where we can specifically check the fraction of apps coming from new vendors, those coming from vendors previously authorized by the user, and those from vendors previously authorized by collaborators. 
		
		\subsubsection{Results for Individuals' Networks}
		Based on these metrics we start by analyzing the results for the individuals' networks, where we observe the following:
		
		\tparagraph{Curtailed growth of privacy loss}
		From Figures~\ref{fig:simulation_groups_inf} and~\ref{fig:simulation_groups_auth}, we notice that the growth of the privacy loss is visibly curtailed in the cases of \EHB\ and \FA\ groups compared to the baseline \EBL\ group. This significant divergence demonstrates the efficacy of our \HB privacy indicators.
		
		\tparagraph{Impact of the network effect}
		Looking into the ratios in Figures~\ref{fig:ratio_inf} and~\ref{fig:ratio_auth}, we see that the privacy loss  in the \EHB\ group has dropped by 41\% in the inflated network and by 28\% in the authors-based network (both with respect to the baseline). In the \FA\ group, where users always optimize their privacy, the privacy loss has dropped by 70\% in the inflated network and by 40\% in the authors-based network. This higher impact in the case of the inflated network is due to the fact that it is a connected graph, unlike the authors-based network, which is composed of smaller connected components. Nevertheless, we can state that, although our privacy indicators have a larger effect on highly connected networks, they are still significantly effective in less connected networks, like the authors-based dataset.

		\tparagraph{Importance of accounting for collaborators' decisions}
		To dive further into events that lead to the observed privacy loss patterns, we look into Figures~\ref{fig:events_inf} and~\ref{fig:events_auth}. First, we observe that users in the \EBL\ group are mainly installing new apps from vendors that had no previous access to their data. This is reflected in the almost linear increase of privacy loss in Figures~\ref{fig:simulation_groups_inf} and~\ref{fig:simulation_groups_auth}. Second, we observe that, in the case of the inflated network, users have been frequently installing apps from vendors with existing access through their collaborators. In fact, as apparent in Figure~\ref{fig:events_inf}, this event outnumbers the event of installing from a new vendor. Third, the number of installations from collaborators' vendors is also significant in the case of the authors-based dataset. While it does not outnumber the installations from new vendors (due to the low-graph connectivity), this is still enough to lead to 28\% and 40\% decrease in the privacy loss in the \EHB\ and the \FA\ groups respectively. Finally, we note that, although the users are more frequently encountering vendors authorized by their collaborators than by themselves, the latter event is still significantly impacting the results. This is because users still incur an incremental privacy loss with vendors authorized by their collaborators while this loss is zero with vendors they have previously authorized. Accordingly, the obtained optimizations are a result of users' accounting for their own and for others' decisions.

\subsubsection{Results for the Teams' Network}
\label{sec:teams}

We now discuss the results for the case of the collaboration network where users work in teams and aim to protect the privacy of the team's data. We observe the following, based on Figure~\ref{fig:sim_team}:
		
		\tparagraph{Inherent usage of similar apps}
		From Figure~\ref{fig:events_team}, it is clear that the dominant event is that of users installing apps which have been authorized by other team members before. This is even in the case of the baseline group (\EBL), which was not the case in the individuals' networks. We justify that by the fact that we selected apps at each simulation step to match their realistic installation frequencies. In practice, apps' installation counts follow a long-tail distribution, and users tend to mostly install a limited set of apps. That is why team members will naturally tend to install a set of similar apps. 
		
		\tparagraph{Curtailed growth of privacy loss}
		Still, we observe that the trend of slower growth of privacy loss also applies in the case of teams (Figure~\ref{fig:simulation_groups_team}). As we also observe in Figure~\ref{fig:ratio_team}, the privacy loss has decreased by 23\% for the \EHB\ group and by 45\% for the \FA\ group, both with respect to the baseline group. This implies that there is an ample room for privacy optimization in teams too.
		
		\tparagraph{Effect due to internal collaborators}
		We finally observe that the privacy loss decrease was achieved via decisions taken by each team's members independently, without relying on other teams' decisions. This highlights the fact that \HB privacy indicators can still be effective even when users do not account for others' decisions. Obviously, taking the external members' decisions into account can lead to further optimizations.\\

		In sum, our simulations {provide further evidence of the efficacy of using History-based privacy indicators in a large network of collaborators.}
		It is worth noting too that, although users in our study were following the \EHB\ decision model, we believe that, {in an actual deployment of such indicators, the model will move closer to the \FA\ model}. This is because users are more protective when their personal data is at risk than when they are put in a role playing scenario about fictitious data. Moreover, users in our study were exposed to this indicator for the first time. When users are educated more about this feature, they might be more likely to take advantage of it.

		\section{Related Work}
		\label{sec:related}
		\subsection{Interdependent Privacy}
		The problem of interdependent privacy has been tackled before in the context of social apps. The main approaches were high-level game-theoretic or economic modeling. 
		In~\cite{biczok2013interdependent}, the authors introduced the concept of interdependent privacy and modeled its impact via a game theoretic, (2-player, 1-app) model. The work by Pu and Grossklags~\cite{pu2014economic} presented a more elaborate economic model that additionally accounts for the interplay among various social network parameters. They showed that app rankings do not accurately reflect the level of interdependent privacy harm the app can cause and that even rational users who consider their friends' well-being might adopt apps with invasive privacy practices. 
		Evidently, these results do not apply in the cloud apps case, where \textit{all} apps have the potential to inflict interdependent privacy harm. 
		
        A later work by Pu and Grossklags~\cite{PuG15} used a conjoint study approach to quantify the monetary value which individuals associate with their friends' personal data. They found that individuals place a significantly higher value on their own personal information than their friends' personal information. This further supports our assumption of self-interested users in this work. 
		The same authors also built on a user survey in~\cite{PuG16} to assess the factors affecting users' own privacy concerns as well as friends' privacy concerns in the context of social app adoption. In particular, they found evidence of negative association between past privacy invasion experiences and the trust in 3rd party apps handling of their own data. They also found partial support for a positive effect of privacy knowledge on concerns for users' own privacy and their friends' privacy. 
        
Other works have also investigated the issue of interdependent privacy in the context of location privacy~\cite{olteanu2016quantifying} and genomic privacy~\cite{humbert2017quantifying}.
In this work, we are focused on quantifying the interdependence of privacy in the context of cloud apps before addressing it from a usable privacy perspective, thus bridging the gap between the theoretical studies and the end-user needs.

		\subsection{Apps Privacy Indicators}
		Our previous work~\cite{Harkous16} was the first to study the privacy of 3rd party cloud apps and to expose that almost two thirds of those apps are over-privileged. In that work, we introduced a novel privacy indicator for deterring users from installing over-privileged apps by showing them Far-reaching insights that apps can needlessly infer from their data (\eg top topics, faces, or locations of interest). In the context of Android apps, Kelly et al., showed that, by adding a set of privacy facts about an app, users will be more likely to choose apps with fewer permissions~\cite{Kelley:2013}. Harbach et al., tackled the same problem but presented users with random examples from their data (\eg pictures, contacts, etc.)~\cite{Harbach:2014}. 
        
		Almuhimedi et al. showed the effectiveness of privacy nudges, which regularly alert users about sensitive data collected by their apps, in encouraging users to review and adjust their permission~\cite{Almuhimedi:2015}.
		All these works, however, tackle the problem of over-privileged apps and try to lead the user into either avoiding them or adjusting their permissions whenever possible. Our current work helps users improve their privacy by reducing the vendors with access to their data, even if the functionality delivered by the vendor abides by the least-privilege principle. Hence, it complements these approaches and can be deployed alongside any of them.

		\section{Conclusion}
		\label{sec:conclusion}
		The findings in this work are the first to concretely delineate the various aspects of interdependent privacy in 3PC apps. One of the major outcomes is that a user's collaborators can be much more detrimental to her privacy than her own decisions. Consequently, accounting for collaborators' decisions should be a key component of future privacy indicators in 3rd party cloud apps. 
		We have shown the impact of History-based Insights as a privacy enhancing technology in this context, especially that, based on our user study, users are less likely to account for previous decisions on their own.
		Our privacy indicators would optimally be implemented by the CSPs themselves as they control the authorization interface and the application stores. The indicators can also be realized by third party privacy providers with access to users' data. 
		Our approach can also be easily mapped to other ecosystems. In the mobile apps' scenario, it can enable users to reduce the number of vendors with access to her contacts. It can also be extended to the case where the goal is protection against 4th parties (e.g., ad providers and data brokers). There, the user can account for data previously held by a 4th party with which the app vendor cooperates. 
		Finally, due to their usability and effectiveness, we envision History-based Insights as an important technique within the movement from static privacy indicators towards dynamic privacy assistants that lead users to data-driven privacy decisions.

	\begingroup
	\raggedright
		\bibliographystyle{abbrv}
		{
			\bibliography{codaspy_ref}}
		\endgroup
		
			\section*{Acknowledgments}
			We would like to thank Deniz Taneli and Nicolas Hubacher for their help in exploratory work that led to this paper. We also thank Rameez Rahman for the helpful discussions and the anonymous reviewers for their valuable feedback. The research leading to these results has received funding from the EU in the context of the project \textit{CloudSpaces}: Open Service Platform for the Next Generation of Personal clouds (FP7-317555).

		\section{Proof of Optimal User Strategy}
		\label{sec:optimal}
		In this section, we complement Section~\ref{sec:HBdesign} by providing a proof the optimal user strategy for minimizing the privacy risk, given our assumptions. We follow the notation introduced in Section~\ref{sec:models}. Let us consider that each 3PC app vendor has a probability $p$ of exposing users' data. As we do not assume that users are provided with a per-vendor risk estimation utility, we set this probability to be the same for all vendors. 
		In general, at a time $t$, a user $u$ would have exposed her data to a set $V$ of vendors, such that each vendor $v$ has access to a fraction $f_{u,v}(t) = \frac{|F_{u,v}(t)|}{|F_u|}$ of the files. Without loss of generality, we will consider henceforth that the user has an all-files privacy goal (cf. Section~\ref{sec:userModel}). However, the same reasoning applies in the case of a per-type privacy goal. In that case, we simply replace ``files'' by ``files of a specific type'' (e.g. photos, documents). We will also be assuming that the users themselves are the data subjects (\ie we consider individual-level subjects).

		For a vendor $v$, we quantify the user's privacy risk magnitude as $p*f_{u,v}(t)$, \ie the fraction of user files possessed by the vendor multiplied by the probability that the vendor exposes the user's files. This vendor could have obtained access due to app installations by the user herself or by her collaborators. 
		A user's privacy risk magnitude at time $t$ can thus be defined as the sum of the risk magnitude across vendors in $V$:  $\mbox{Risk}(t) = \sum_{v\in V} p*f_{u,v}(t)$.
		
		When a user installs an app from a vendor $\hat{v}$ at time $t+1$, the vendor gets access to the whole set of user's files. Hence, the risk magnitude is increased by  $p*(1- f_{u,\hat{v}}(t))$. Given that $p$ is constant, the risk magnitude can be minimized by choosing $\hat{v}$, such that $\hat{v}=\argmax_v f_{u,v}(t)$ (which can also be written as   $\hat{v}=\argmax_v \VFC_u(\{ v \},t)$).
Hence, the optimal, greedy strategy to minimize the risk is to select the vendor that already has the largest fraction of user files, thus minimizing $p*(1- f_{u,\hat{v}}(t))$. We call this strategy: ``History-based decisions''.

	\end{document}